\documentclass[conference]{IEEEtran}
\IEEEoverridecommandlockouts
\usepackage[usenames,table,dvipsnames]{xcolor}
\usepackage[notes=xxx]{dtrt}

\usepackage[
style=ieee,
giveninits,
backend=bibtex8
]{biblatex}
\bibliography{IEEEabrv,sumcheck_refs}

\usepackage{amsthm}

\usepackage[operators]{cryptocode}
\usepackage{mathptmx}
\usepackage{stmaryrd}   
\usepackage{microtype}
\usepackage{tikz}
\usetikzlibrary{matrix}
\usetikzlibrary{shapes,arrows}
\usepackage{graphicx}
\usepackage{verbatim}
\usepackage{array}
\usepackage{multirow}
\usepackage{latexsym}
\usepackage{paralist}
\usepackage{enumitem}
\setlist[itemize]{leftmargin=*}
\setlist[enumerate]{leftmargin=*}
\usepackage[capitalise]{cleveref}
\crefname{step}{Step}{Steps}
\usepackage{dsfont}
\usepackage{url}
\usepackage{cleveref}
\usepackage{braket}
\usepackage{mathrsfs}
\usepackage{color}
\usepackage{soul}
\usepackage{mdframed}
\usepackage{complexity}
\usepackage{mathtools}
\usepackage{bbm}
\usepackage{setspace}
\usepackage{tabu}
\usepackage{multirow}
\usepackage{adjustbox}
\usepackage{booktabs}
\usepackage{float}
\usepackage{hhline}
\usepackage{tcolorbox}

\usepackage{biblatex}
\usepackage{hyperref}
\usepackage{xurl}
\hypersetup{breaklinks=true}

\crefname{appsec}{appendix}{appendices}
\Crefname{appsec}{Appendix}{Appendices}

\usepackage[margin=4mm,small,labelfont=bf]{caption}

\def\BibTeX{{\rm B\kern-.05em{\sc i\kern-.025em b}\kern-.08em
    T\kern-.1667em\lower.7ex\hbox{E}\kern-.125emX}}

\newtheorem{theorem}{Theorem}[section]

\newtheorem{definition}[theorem]{Definition}

\newtheorem{protocol}[theorem]{Protocol}

\allowdisplaybreaks
\newcommand{\FormatAuthor}[3]{
	\begin{tabular}{c}
		#1 \\ {\small\texttt{#2}} \\ {\small #3}
	\end{tabular}
}

\newcommand{\defemph}[1]{\textbf{\emph{#1}}}
\newcommand{\defeq}{:=}

\newcommand{\Alt}[1]{{#1}'}

\newcommand{\N}{\mathbb{N}}

\newcommand{\Bits}{\{0,1\}}
\newcommand{\Bitstring}{\Bits^*}
\newcommand{\F}{\mathbb{F}}
\newcommand{\UnivA}{\mathbb{A}}

\newcommand{\mysigma}{\textstyle\scalebox{0.9}{$\sigma$}}
\newcommand{\myalpha}{\textstyle\scalebox{0.9}{$\alpha$}}
\newcommand{\myrho}{\textstyle\scalebox{0.9}{$\rho$}}

\newcommand{\Prob}{\textsf{Pr}}

\newcommand{\Language}{\mathcal{L}}

\newcommand{\Instance}{x}

\newcommand{\Pro}{\textbf{P}}
\newcommand{\Ver}{\textbf{V}}
\newcommand{\Interact}[2]{\langle#1,#2\rangle}

\newcommand{\Mal}[1]{\hat{#1}}

\newcommand{\CheckEq}{\stackrel{?}{=}}

\newcommand{\VerRandomness}{\isa{r}}
\newcommand{\Message}{a}

\newcommand{\State}[2]{s_{#1}^{#2}}
\newcommand{\NewState}[1]{s'_{#1}}

\newcommand{\NumRounds}{k}

\newcommand{\SoundnessError}{s}
\newcommand{\CompletenessError}{c}

\newcommand{\SPSymbol}{{\scriptscriptstyle\mathrm{SP}}}

\newcommand{\SPLanguage}{\Language_\SPSymbol}

\newcommand{\SPSubset}{\isa{H}}
\newcommand{\SPArity}{\isa{n}}
\newcommand{\SPTarget}{\isa{v}}
\newcommand{\SPPolynomial}{\isa{p}}
\newcommand{\SPProver}{\isa{pr}}
\newcommand{\SPDegree}{\isa{d}}
\newcommand{\SPState}{\isa{s}}
\newcommand{\SPRandomness}{\isa{r}}

\newcommand{\NewSPTarget}{\SPTarget'}
\newcommand{\NewSPPolynomial}{\SPPolynomial'}
\newcommand{\NewSPState}{\SPState'}
\newcommand{\NewSPRandomness}{\SPRandomness'}

\newcommand{\SPInstanceTuple}{(\F,\SPSubset,\SPPolynomial,\SPTarget)}
\newcommand{\NewSPInstanceTuple}{(\F,\SPSubset,\NewSPPolynomial,\NewSPTarget)}
\newcommand{\SPInstanceTupleIsa}{(\SPSubset,\SPPolynomial,\SPTarget)}

\newcommand{\SPInstanceShort}{\mathit{sc}}
\newcommand{\NewSPInstanceShort}{\SPInstanceShort'}

\newcommand{\XVar}{X}
\newcommand{\SumVal}{x}

\newcommand{\SPMessagePolynomial}{\isa{q}}

\newcommand{\Accept}{\textsf{acc}}

\newcommand{\PolyEq}{\SPMessagePolynomial = \Mal{\SPMessagePolynomial}}
\newcommand{\PolyNeq}{\SPMessagePolynomial \neq \Mal{\SPMessagePolynomial}}

\newcommand{\EvalEq}{\SPMessagePolynomial(\VerRandomness_1) = \Mal{\SPMessagePolynomial}(\VerRandomness_1)}
\newcommand{\EvalNeq}{\SPMessagePolynomial(\VerRandomness_1) \neq \Mal{\SPMessagePolynomial}(\VerRandomness_1)}

\DeclareTextFontCommand{\isakw}{\rmfamily\bfseries}  
\newcommand{\isaco}[1]{\mathsf{#1}}                  
\DeclareTextFontCommand{\isaid}{\rmfamily\itshape}   

\newcommand{\isabf}{\isakw}
\newcommand{\isa}{\isaid}

\newcommand{\isalemma}{\isakw{lemma }}
\newcommand{\isatheorem}{\isakw{theorem }}

\newcommand{\isalocale}{\isakw{locale }}
\newcommand{\isafixes}{\isakw{fixes }}
\newcommand{\isaassumes}{\isakw{assumes }}
\newcommand{\isashows}{\isakw{shows }}
\newcommand{\isaand}{\isakw{ and }}

\newcommand{\isatypesynonym}{\isakw{type\_synonym }}
\newcommand{\isatypedef}{\isakw{typedef }}

\newcommand{\isadefinition}{\isakw{definition }}
\newcommand{\isaliftdefinition}{\isakw{lift\_definition }}
\newcommand{\isafun}{\isakw{fun }}
\newcommand{\isawhere}{\isakw{ where }}
\newcommand{\isais}{\isakw{is }}

\newcommand{\isabegin}{\isakw{begin}}
\newcommand{\isaend}{\isakw{end}}

\newcommand{\isalet}{\isakw{let }}
\newcommand{\isain}{\isakw{ in}}
\newcommand{\isaif}{\isakw{if }}
\newcommand{\isathen}{\isakw{ then }}
\newcommand{\isaelse}{\isakw{ else }}

\newcommand{\fun}{\Rightarrow}

\newcommand{\card}[1]{\lvert #1 \rvert}

\newcommand{\isand}{\:\land\:}

\newcommand{\verchecks}{C_{\isa{ver}}}    
\newcommand{\rvcheck}{C_{\isa{rec}}}
\newcommand{\failure}{F}

\newcommand{\Probleft}{\Prob_{\isa{left}}}
\newcommand{\Probright}{\Prob_{\isa{right}}}

\newcommand{\fst}{\isaco{fst}}
\newcommand{\snd}{\isaco{snd}}

\newcommand{\cons}{\mathrel{\#}}

\newcommand{\mapfun}{\isaco{map}}
\newcommand{\distinct}{\isaco{distinct}}
\newcommand{\setfun}{\isaco{set}}

\newcommand{\dom}{\isaco{dom}}
\newcommand{\mapof}{\isaco{map\_of}}
\newcommand{\mapupdate}{\mathrel{++}}
\newcommand{\mapempty}{\emptyset}      

\newcommand{\prove}{\isa{prove}}

\newcommand{\substs}{\isaco{substs}}
\newcommand{\tuples}{\isaco{tuples}}

\newcommand{\vars}{\isa{vars}}
\renewcommand{\deg}{\isa{deg}}    
\newcommand{\eval}{\isa{eval}}
\newcommand{\inst}{\isa{inst}}

\newcommand{\varsco}{\isaco{vars}}
\newcommand{\degco}{\isaco{total\_degree}}
\newcommand{\evalco}{\isaco{eval}}
\newcommand{\instco}{\isaco{inst}}

\newcommand{\sumcheck}{\isaco{sumcheck}}
\newcommand{\honestprover}{\isaco{honest\_prover}}

\renewcommand{\epsilon}{\ensuremath\varepsilon}

\renewcommand{\phi}{\ensuremath{\varphi}}

\begin{document}

\title{Formal Verification of the Sumcheck Protocol}

\author{
    \begin{tabular}[h!]{ccc}
        \FormatAuthor{Azucena Garvía Bosshard}{azucena.garvia-bosshard@alumni.ethz.ch}{Independent Researcher${}^\dagger$\thanks{$\dagger$ Work done while at ETH Zurich}}
        &
        \FormatAuthor{Jonathan Bootle}{jbt@zurich.ibm.com}{IBM Research Europe -- Zurich}
        &
        \FormatAuthor{Christoph Sprenger}{sprenger@inf.ethz.ch}{ETH Zurich}
    \end{tabular}
}
\date{\today}

\maketitle

\begin{abstract} 
The sumcheck protocol, introduced in 1992, is an interactive proof which is a key component of many probabilistic proof systems in computational complexity theory and cryptography, some of which have been deployed. However, none of these proof systems based on the sumcheck protocol enjoy a formally-verified security analysis.
In this paper, we make progress in this direction by providing a formally verified security analysis of the sumcheck protocol using the interactive theorem prover Isabelle/HOL. We follow a general and modular approach.

First, we give a general formalization of public-coin interactive proofs. We then define a \emph{generalized sumcheck protocol} for which we axiomatize the underlying mathematical structure and we establish its soundness and completeness. Finally, we prove that these axioms hold for multivariate polynomials, the original setting of the sumcheck protocol.
Our modular analysis facilitates formal verification of sumcheck instances based on different mathematical structures with little effort, by simply proving that these structures satisfy the axioms.
Moreover, the analysis supports the development and formal verification of future cryptographic protocols using the sumcheck protocol as a building block.
\end{abstract}


\section{Introduction}

Probabilistic proof systems are protocols in which a powerful but untrusted prover can convince a verifier that a given statement is true. They differ from traditional mathematical proofs by incorporating interaction, randomness and cryptographic assumptions. As a result, they offer benefits such as fast verification and zero-knowledge properties.

Early research into interactive proofs \cite{introductionIP,eqintroductionIP} produced complexity theoretic results such as $\mathsf{IP}=\mathsf{PSPACE}$.
Today, probabilistic proof systems enjoy a high degree of adoption and are increasingly deployed in applications such as verifiable computation, electronic voting, and cryptocurrencies.

\parhead{Errors in analysis}

Unfortunately, the designs and deployments of probabilistic proof systems often contain errors:
\begin{itemize}
    \item The analysis of the proof system \cite{FujisakiO97} for integer commitments contained errors, later addressed in \cite{DamgardF02};
    \item The public setup information of the proof system \cite{Ben-SassonCTV14} contained extra information which made the proof system insecure. This affected the ZCash cryptocurrency \cite{SWB19};
    \item A significant gap in the security proof of \cite{darksnark} was found and fixed using three different methods in \cite{darkfix1,darkfix2,BunzF22}.
    \item The proof system described in \cite{bootleZK} states a cheating probability lower than its actual value. A proper analysis was given in \cite{bootlefix1, bootlefix2}.
    \item The description of the Fiat-Shamir transformation in \cite{BuenzBBPWM18} neglects to hash certain values, which leads to the vulnerability described in \cite{Miller22}. This affected multiple implementations of proof systems \cite{ZenGo,SecBit,Dusk,Iden3,gnark}.
\end{itemize}

Other researchers have raised concerns about errors in cryptographic analyses in general~\cite{bellare2004code,CAC}. As probabilistic proofs have become increasingly popular, their security analyses demand greater care and scrutiny.

\parhead{Formal verification of probabilistic proof systems}

In order to remedy these issues, researchers are starting to turn to machine-checkable security analyses using formal verification techniques.
Existing formal verification work falls into two categories: formal verification of non-interactive proof systems, such as that of \cite{Groth16} by \cite{BM23,BoltonBailey}, and constant-round interactive proof systems, including Sigma protocols \cite{FU22}, protocols using the ``MPC-in-the-head'' paradigm of \cite{IshaiKOS09} in \cite{MPCformalization,ABEGPP21}, and particular signature schemes \cite{AFOU23,dilithium_formalization}.

However, despite this growing body of work, interactive proof systems with a non-constant number of rounds have not yet been addressed. This category contains a large number of protocols, including proof systems based on the GKR protocol \cite{ref1auro} for uniform circuits, proof systems for NP \cite{spartan}, ``split-and-fold'' proof systems such as \cite{BootleCCGP16,BuenzBBPWM18}, as well as protocols with optimal prover complexity \cite{tensor1,tensor3,RonZewiR22,HolmgrenR22} and interesting streaming properties \cite{streamingIPs,BlockHRRS20,darkfix1,gemini}. Note that while e.g. \cite{FU22} does consider sequential composition of Sigma protocols, leading to a multi-round protocol, many of the protocols cited above have a significantly more complicated recursive structure.

In fact, the multi-round protocols cited above are all based on a single protocol: the sumcheck protocol. Each protocol makes direct use of the sumcheck protocol, or its algebraic \cite{scarguments} or combinatorial \cite{tensorsumcheck} generalizations, as a subroutine. Moreover, the streaming properties of \cite{BlockHRRS20,darkfix1,gemini} are directly linked to those of the sumcheck protocol.

\parhead{The sumcheck protocol}

As the name suggests, the goal of the sumcheck protocol is to check that the evaluation of a sum corresponds to a certain value. More specifically, the verifier $\Ver$'s objective is to use the prover $\Pro$ to confirm that 
\begin{equation} \label{eq:sumcheck-problem}
    \SPTarget = \sum_{(\SumVal_1,\ldots,\SumVal_\SPArity) \in \SPSubset^\SPArity} \SPPolynomial(\SumVal_1,\ldots,\SumVal_\SPArity)
    \enspace,
\end{equation}
where $\SPSubset$ is a subset of some field $\F$, $\SPPolynomial$ is a polynomial of arity $\SPArity$ over $\F$, and $\SPTarget \in \F$. The protocol has $\SPArity$ rounds of interaction between the prover and the verifier, and a recursive structure whereby statements about polynomials with arity $\SPArity$ are reduced to statements about polynomials with smaller arity, repeatedly, until arity $0$ is reached.

As well as being used in many practical probabilistic proof systems, the sumcheck protocol and its generalizations are also a core component of many complexity-theoretic results, like $\mathsf{coNP} \subseteq \mathsf{IP}$, $\mathsf{PSPACE} = \mathsf{IP}$ \cite{IPisPSPACE,shen1992ip}, and results on $\mathsf{PPAD}$-hardness \cite{Jawale2020SNARGsFB,ChoudhuriHKPRR19,KalaiLV23}. It is also one of few interactive proof systems that can be made non-interactive via the Fiat-Shamir transformation without relying on heuristics like the random oracle model \cite{CanettiCHLRR18}.

Given the versatility of the sumcheck protocol, it is clear that its formal analysis is highly desirable and constitutes the first step towards analysing a great many probabilistic proofs with unique and interesting properties.

\subsection{Our Work}

We provide a formally verified security analysis of the sumcheck protocol \cite{introSumcheck}.  
We start with generic definitions of (recursive, multi-round) public-coin interactive proofs and the associated soundness and completeness properties, which we refer to as security properties. Our definition is parameterized by prover and verifier functions and reduces the size of the problem instance in each recursive call.
We then formalize a \emph{generalized sumcheck protocol} that abstracts from the concrete mathematical structure of polynomials and we provide machine-checkable mathematical proofs of its security properties in Isabelle/HOL. Our proofs thus cover an entire protocol family of which the concrete sumcheck protocol is a special case. We also show that the protocol and its properties are an instance of our generic definitions for public-coin interactive proofs. 
More precisely, our analysis proceeds in two steps.

In the first step, we axiomatize the key properties of polynomials required by the protocol, formalize the generalized sumcheck protocol based on these axioms, and prove its security. We have made an effort to make the the axioms as weak as possible to ensure their wide applicability. For example, we make only minimal assumptions about the polynomials' algebraic structure, namely, that they form a commutative monoid over a finite support set.
For the inductive proofs to succeed, the soundness and completeness theorems required several generalizations of the statements usually found in less formal accounts. For example, in the generalized soundness statement, 
the assumptions about the polynomial's arity and degree must be generalized to ensure that the induction hypothesis is applicable. 

In a second step, we show that concrete multivariate polynomials satisfy our axioms. We do this based on an existing (concrete) formalization of multivariate polynomials in Isabelle~\cite{Polynomials-AFP}, which we extend with additional definitions and results, e.g., about polynomial instantiation and the number of a polynomial's roots. This establishes the axioms' consistency and specializes the security proofs to the concrete protocol instance. 

\subsection{Our Contributions}

Our main contributions are as follows. 
\begin{itemize}
\item We define and formalize a generalized version of the sumcheck protocol and prove its soundness and completeness. We then instantiate this protocol with multivariate polynomials. To the best of our knowledge, this is the first machine-checkable security analysis for probabilistic proof systems with an input-size-dependent number of rounds. 

\item Our general and modular analysis has several benefits. 
First, it simplifies proofs by decoupling the reasoning about the protocol from reasoning about concrete polynomial representations. 
Second, our general results can be used to obtain security proofs of sumcheck protocol variants~\cite{CCKP19,scarguments} relying on different but similar mathematical structures, such as, for example,
rings or modules instead of finite fields. This can be achieved with minimal effort by proving that these structures satisfy the required axioms, with some minor changes.
Finally, our work paves the way for the future analysis of the many aforementioned protocols which use the sumcheck protocol as a subroutine.
\end{itemize}

\subsection{Related Work}

Bailey and Miller \cite{BM23,BoltonBailey} formalized the security analyses of various non-interactive proof systems in the Lean~3 interactive theorem prover. They analyze \cite{GennaroGP013,ParnoHG013,BagheryKSV21,Lipmaa19,babysnark}, and the construction of \cite{Groth16} which has extremely low communication complexity and is widely deployed. All of these proof systems are built from cryptographic pairings and have very similar properties. Their security analysis takes place in a strong idealized model called the ``algebraic group model'', which simplifies both pen-and-paper and machine security analysis. The analysis focuses on verifying equality between the coefficients of various multivariate polynomial expressions.

Work on constant-round proof systems includes analyses of Sigma protocols and their composition properties in \cite{FU22} and protocols using the ``MPC-in-the-head'' paradigm of \cite{IshaiKOS09} in \cite{MPCformalization,ABEGPP21}. Many signature schemes are also based on constant-round proof systems, and machine-formalized signature schemes and implementations of this type include Schnorr \cite{AFOU23} and Dilithium \cite{dilithium_formalization} signatures. All of these formalizations use EasyCrypt. The analysis of Sigma protocols is made more challenging through the use of computational assumptions and security definitions involving computationally bounded adversaries, and rewinding algorithms, which we do not consider here. Note that \cite{FU22} does consider the sequential composition of Sigma protocols, leading to the formal analysis of a multi-round protocol. However, a composed Sigma protocol treats the same instance in each repetition of the Sigma protocol, whereas the sumcheck protocol reduces to a new instance in each round of the protocol, in a way that depends on the previous messages in the protocol.
Both Sigma protocols and commitment schemes were formalized in~\cite{cryptholsigmacommitments} using the CryptHOL framework based on Isabelle/HOL.

A recent Isabelle formalization of the Schwartz-Zippel lemma \cite{SchwartzZippel-AFP} also involves formalizing various probabilistic and inductive statements about multivariate polynomials similar to some lemmas we prove as part of our soundness proof.

\subsection{Paper Outline and Supplementary Material}

In \Cref{sec:preliminaries}, we introduce interactive proofs, the sumcheck protocol, Isabelle/HOL, and notation. We then formalize public-coin interactive proofs in Isabelle (\Cref{sec:formalizing-public-coin-proofs}). In \Cref{sec:generalized-sumcheck}, we present our formalization of a generalized  sumcheck protocol and in \Cref{sec:security-properties} proofs of its security properties. In \Cref{sec:polynomial-instantiation}, we instantiate the generalized sumcheck protocol to the original one using polynomials.

The complete Isabelle/HOL sources for the development in this paper are available online~\cite{Sumcheck-AFP-2024}.

\section{Preliminaries}
\label{sec:preliminaries}

\subsection{Interactive Proofs}
\label{sec:ip}
Interactive proofs between a \emph{prover} algorithm $\Pro$ and a \emph{verifier} algorithm $\Ver$ were first introduced by Goldwasser, Micali and Rackoff in \cite{introductionIP}. In this section, we give formal definitions of interactive proofs in terms of a sequence of function executions between $\Pro$ and $\Ver$. We start by defining interactions.
\begin{definition}[Interactions]
\label{def:classic_stateful_interaction}
Let $\Pro, \Ver : \Bitstring \rightarrow \Bitstring$ and $\NumRounds \colon \Bitstring \rightarrow \N$ be functions.
We define a \defemph{$\NumRounds(\Instance)$-round interaction} between $\Pro$ and $\Ver$ on input $\Instance \in \Bitstring$ with verifier randomness $\VerRandomness$ as the following sequence of function outputs:
\begin{align*}
    \Message_1, \State{\Pro}{1} &= \Pro(\State{\Pro}{0})
    \enspace,\\
    \Message_2, \State{\Ver}{1} &= \Ver(\State{\Ver}{0}, \Message_1)
    \enspace,\\
    \Message_3, \State{\Pro}{2}  &= \Pro(\State{\Pro}{1}, \Message_2)
    \enspace,\\
    &\vdots\\
    \Message_{2\NumRounds(\Instance)}, \State{\Ver}{\NumRounds(\Instance)} &= \Ver(\State{\Ver}{\NumRounds(\Instance)-1}, \Message_{2\NumRounds(\Instance)-1})
    \enspace.
\end{align*}
Here, $\Message_{2i-1}$ are messages from prover to verifier, while $\Message_{2i}$ are messages from verifier to prover. Further, $\State{\Pro}{i}$ and $\State{\Ver}{i}$ are the states of the prover and verifier respectively at step $i$. The inputs and outputs are concatenated into single bitstrings using commas, and we assume suitable encodings of input values that allow tuples to be parsed unambiguously. The initial prover state $\State{\Pro}{0}$ is set to $\Instance$ and the initial verifier state $\State{\Ver}{0}$ is set to $(\Instance,\VerRandomness)$. 
The outcome of the interaction is denoted by $\Interact{\Pro}{\Ver(\VerRandomness)}(\Instance) = \Message_{2\NumRounds(\Instance)}$.
\end{definition}
A round in an interaction consists of a message from the prover to the verifier and the verifier's corresponding response. These are two moves. 
Note that only the verifier has access to the randomness $\VerRandomness$ of the interaction, ensuring that the prover cannot predict the verifier's messages.

Now, we define languages, and interactive proofs with respect to particular languages.

\begin{definition} [Languages]
\label{def:languages}
A \defemph{language} is a set $\Language \subseteq \Bitstring$ of binary strings. An element $\Instance \in \Language$ is referred to as an \defemph{instance}.
\end{definition}
\begin{definition}[Interactive Proof Systems]
\label{def:IPsystems}
Let $0< \SoundnessError < 1 - \CompletenessError < 1$. Let $\NumRounds \colon \Bitstring \rightarrow \N$.
An \defemph{interactive proof system} for a language $\Language$ with completeness error $\CompletenessError$ and soundness error $\SoundnessError$ is a $\NumRounds(\Instance)$-round interaction between $\Pro$ and $\Ver$  satisfying the following conditions:
\begin{itemize}
    \item \textbf{Completeness:} For all $\Instance \in \Language$,
    \begin{equation*}
        \Prob_\VerRandomness [\langle \Pro, \Ver(\VerRandomness) \rangle (\Instance) = 1] \geq 1-\CompletenessError
        \enspace.
    \end{equation*}
    
    \item \textbf{Soundness:} For all $\Instance \notin \Language$ and for all $\Mal{\Pro}$,
    \begin{equation*}
        \Prob_\VerRandomness [\langle \Mal{\Pro}, \Ver(\VerRandomness) \rangle (\Instance) = 1] \leq \SoundnessError 
        \enspace.
    \end{equation*}

    \item The total running time of the verifier $\Ver$ is polynomially bounded in the length of $\Instance\in\Bitstring$.
\end{itemize}
Here, probabilities are taken over the random value $\VerRandomness$ which denotes the randomness used by the verifier in the protocol. 
If $\langle \Pro, \Ver(\VerRandomness) \rangle (\Instance) = 1$ then we say that the verifier \defemph{accepts}, and otherwise, we say that the verifier \defemph{rejects}.
\end{definition}

Intuitively, completeness ensures the honest prover can usually convince the verifier of a true statement, and soundness guarantees a dishonest prover cannot usually convince the verifier of a false statement. The completeness and soundness errors bound the probability that the verifier errs.

\Cref{def:classic_stateful_interaction} is closely related to the definition of interactive Turing machines in \cite{introductionIP}, with the prover and verifier states taking on the role of the work and input tapes in the Turing machine definition. \Cref{def:classic_stateful_interaction} only considers deterministic provers, but this does not alter the class of languages $\Language$ for which an interactive proof exists.

Note that the completeness and soundness errors in \Cref{def:IPsystems} can be generalized to functions $\CompletenessError,\SoundnessError \colon \N \to [0,1]$ such that $\CompletenessError(|\Instance|) < 1 - 2^{\poly (|\Instance|)}$, $\SoundnessError(|\Instance|) > 2^{-\poly(|\Instance|)}$ and $\CompletenessError(|\Instance|) > \SoundnessError(|\Instance|) + \frac{1}{\poly(|\Instance|)}$. The completeness and soundness bounds are then replaced by $\CompletenessError(|\Instance|)$ and $\SoundnessError(|\Instance|)$. This does not change the class of languages $\Language$ for which an interactive proof exists.

Finally, we define public-coin interactive proofs.
\begin{definition}[Public-Coin Proofs]
\label{def:public-coin-proofs}
    We say that an interactive proof system is \defemph{public coin} if each of the verifier's messages is chosen uniformly at random from some set.
\end{definition}

\subsection{The Sumcheck Protocol}
\label{sec:sumcheck}

In this section, we introduce the sumcheck protocol~\cite{introSumcheck}, a public-coin interactive proof for the sumcheck problem, which is defined by the following language. We assume suitable methods of encoding tuples and their elements into binary strings.
\begin{definition}
\label{def:sp}
    Let $\SPLanguage$ be the set of tuples $\SPInstanceTuple$ such that $\F$ is a finite field, $\SPSubset \subseteq \F$, $\SPPolynomial \colon \F^\SPArity \to \F$ is a polynomial of arity $n$, and $v \in \F$, satisfying
    \begin{equation*}
        \SPTarget = \sum_{(\SumVal_1,\ldots,\SumVal_\SPArity) \in \SPSubset^\SPArity} \SPPolynomial(\SumVal_1, \SumVal_2, \ldots, \SumVal_\SPArity)
        \enspace.
    \end{equation*}
\end{definition}

This is the classic definition of the sumcheck problem with polynomials over finite fields. Both the language $\SPLanguage$ and \Cref{prot:recursive_protocol} below generalize to other structures satisfying suitable properties, such as polynomials over rings and modules. This will be discussed further in \Cref{sec:generalized-sumcheck}.

We give a recursive description of the sumcheck protocol in \Cref{prot:recursive_protocol}.
\begin{protocol}[Sumcheck Protocol]
\label{prot:recursive_protocol}
    The prover $\Pro$ takes as input a state $\State{\Pro}{}$ containing the instance $\Instance = \SPInstanceTuple$.
    The verifier $\Ver$ takes as input a state $\State{\Ver}{}$ containing the instance $\Instance$ and randomness $(\VerRandomness_1,\ldots,\VerRandomness_\SPArity)\in\F^\SPArity$, where $\SPArity$ is the arity of $\SPPolynomial$.
    \begin{itemize}
        \item If $\SPPolynomial$ has arity $0$, then $\SPPolynomial$ is a constant polynomial. The prover $\Pro$ does not send a message to the verifier. The verifier $\Ver$ checks whether $\SPPolynomial = \SPTarget$ and accepts the proof if so, rejecting otherwise.
        
        \item Otherwise, the prover $\Pro$ sends $\Ver$ the polynomial
        \begin{equation*}
            \SPMessagePolynomial(\XVar) \defeq \sum_{(\SumVal_2,\ldots,\ldots,\SumVal_\SPArity) \in \SPSubset^{\SPArity-1}} \SPPolynomial( \XVar, \SumVal_{2},\ldots,\SumVal_\SPArity).
        \end{equation*}
        
        \item The verifier $\Ver$ checks that $\SPMessagePolynomial$ is a univariate polynomial of degree at most $\deg(\SPPolynomial)$, and that $\SPTarget \CheckEq \sum_{\SumVal\in \SPSubset} \SPMessagePolynomial(\SumVal)$. 
        If any of the checks fail, $\Ver$ rejects the proof.\footnote{Early rejection by the verifier before the end of the protocol is easily modelled by having the verifier include a special symbol in their state to ensure that they do reject at the end.}   
        Otherwise $\Ver$ sends randomness $\VerRandomness_1$ to $\Pro$.

        \item The verifier $\Ver$ computes a reduced instance $\Alt{\Instance}=(\F,\SPSubset,\Alt{\SPPolynomial},\Alt{\SPTarget})$ where $\Alt{\SPPolynomial}(\XVar_2,\ldots,\XVar_\SPArity) \defeq \SPPolynomial(\VerRandomness_1,\XVar_2,\ldots,\XVar_\SPArity)$ and $\Alt{\SPTarget} \defeq \SPMessagePolynomial(\VerRandomness_1)$, and updates its state to $\NewState{\Ver}\defeq (\Alt{\Instance},(\VerRandomness_2,\ldots,\VerRandomness_\SPArity))$.
        The prover $\Pro$ updates its state to $\NewState{\Pro} \defeq \Alt{\Instance}$.
        The prover $\Pro$ and verifier $\Ver$ continue the protocol on the reduced instance.
    \end{itemize}
\end{protocol}

Note that the arity of the polynomial is reduced in each round and therefore the protocol terminates after at most $\SPArity$ reductions.

The main goal of this paper is to produce a machine-checked formalisation of the following theorem and its proof.

\begin{theorem}
\label{thm:paper-sumcheck-theorem}
The sumcheck protocol defines a public-coin interactive proof for the language $\Language_\SPSymbol$ with null completeness error and soundness error $\SPArity\cdot  \deg(\SPPolynomial) / |\F|$, where $\SPPolynomial$ is the instance polynomial and $\SPArity$ its arity.
\end{theorem}
The proof of this theorem is given in \Cref{app:properties}.

Here, and throughout the paper, $\deg$ refers to the total degree of a multivariate polynomial. The verifier check that $\SPMessagePolynomial$ has degree at most $\deg(\SPPolynomial)$ is required for the soundness proof of the sumcheck protocol, which relies on the fact that the number a polynomial's roots is bounded by its degree. In fact, $\deg(\SPPolynomial)$ could be defined as $\max_{i \in \{1,\ldots,\SPArity\}} \deg_{i}(\SPPolynomial)$, where $\deg_{i}(\SPPolynomial)$ is the degree of $\SPPolynomial$ in variable $x_i$, which gives a tighter bound on the soundness error.

\subsection{Isabelle/HOL and Notation}
\label{subsec:notation}

Isabelle is a generic interactive proof assistant that can be instantiated to different object logics. We use its higher-order logic instance, Isabelle/HOL~\cite{NipkowPW02}, which uses the simply typed $\lambda$-calculus as its underlying specification language. On top of that it offers various convenient high-level specification mechanisms, e.g., for datatypes, inductive definitions, and recursive function definitions. Isabelle/HOL comes with an extensive library of concepts and theorems (e.g., about algebra, analysis, probability theory). Isabelle is a foundational prover, i.e., all logical inferences are checked by a small and well-tested logical kernel, and therefore it provides very strong soundness guarantees. Isabelle offers a wide range of automated proof tools. The main ones are a simplifier for rewriting (equational reasoning), a classical reasoner, and sledgehammer, a binding to various back-end automated theorem provers and SMT solvers. 

An Isabelle/HOL development is organized into a collection of theories, each defining new types and constants and proving theorems about them. We write variables and types in italics (e.g., $\isa{nat}$) and constants in sans serif font (e.g., $\isaco{zip}$). Type variables are marked with a leading quote (e.g., \isa{'a}). We write $t :: T$ to indicate that term~$t$ has type~$T$. Isabelle also supports type classes, which are used to axiomatize types with a certain (e.g., algebraic) structure. The (overloaded) notation $\isa{'a} :: \tau$ indicates that the type $\isa{'a}$ belongs to type class $\tau$. For example,  $\isa{xs} :: (\isa{'a} :: \isa{comm\_monoid})~\isa{list}$ indicates that $\isa{xs}$ is a list of elements of a type $\isa{'a}$ with a commutative monoid structure. 

We extensively use library functions for lists and maps (i.e., partial functions): $x \cons \isa{xs}$ prepends the element $x$ to the list $\isa{xs}$, $\mapfun~ f~ \isa{xs}$ applies the function $f$ to all elements of $\isa{xs}$, $\isaco{zip}~\isa{xs}~\isa{ys}$ turns the given lists into a list of pairs (truncating the longer list, if any), $\isaco{set}~\isa{xs}$ turns $\isa{xs}$ into a set, $\distinct~\isa{xs}$ asserts the uniqueness of all list elements, $\dom~m$ denotes the map $m$'s domain, 
and $m_1 \mapupdate m_2$ overrides the  map $m_1$ with the entries of $m_2$.
We also use some own auxiliary functions: $\tuples~n$ denotes the set of lists of length $n$ and 
$\substs~V~H$ refers to the set of maps from elements of $V$ to elements of $H$, which we use to model substitutions.

We sometimes take some notational liberties for the sake of readability. For example, we write $\Prob_{\isa{r} \:\in\: \isa{S}}[\isa{E} ~ \isa{r}]$ for the Isabelle term $(\isaco{measure\_pmf.prob} ~ (\isaco{pmf\_of\_set} ~ \isa{S}) ~ \{\isa{r}.\; \isa{E} ~ \isa{r}\})$ denoting the probability of the event $E$ under a uniformly random $r$ sampled from the set $S$.

\section{Formalizing Public-Coin Proofs}
\label{sec:formalizing-public-coin-proofs}

\begin{figure}[t]
\begin{align*}
  & \isatypesynonym (\isa{'i}, \isa{'r}, \isa{'a}, \isa{'resp}, \isa{'ps}) ~ \isa{prv} = \\
  & \quad\;  
    \isa{'i} \fun \isa{'a} \fun \isa{'a} ~ \isa{list} \fun \isa{'r} \fun \isa{'ps} \fun \isa{'resp} \times \isa{'ps} \\[1.5ex]
  & \isalocale \isa{public\_coin\_proof} = \\
  & \isafixes \isa{ver}_0 :: \isa{'i} \fun \isa{'vs} \fun \isa{bool} \\
  & \isaand \isa{ver}_1 :: \isa{'i} \fun \isa{'resp} \fun \isa{'r} \fun \\
  & \phantom{\isaand \isa{ver}_1 :: {}} 
    \isa{'a} \fun \isa{'a}~\isa{list} \fun \isa{'vs} \fun \isa{bool} \times \isa{'i} \times \isa{'vs} \\
  & \isabegin \\
  & \isafun \prove :: \isa{'vs} \fun (\isa{'i}, \isa{'r}, \isa{'a}, \isa{'resp}, \isa{'ps})~ \isa{prv} \fun \isa{'ps} \fun \\
  & \phantom{\isafun \prove :: {}} \isa{'i} \fun \isa{'r} \fun (\isa{'a} \times \isa{'r})~ \isa{list} \fun \isa{bool} \isawhere \\
  & \quad \prove ~\isa{vs} ~ \isa{prv} ~ \isa{ps} ~ \isa{I} ~ \isa{r} ~ [] \,\longleftrightarrow\, 
    \isa{ver}_0 ~ \isa{I} ~ \isa{vs} \\
  & \quad \prove ~ \isa{vs} ~\isa{prv} ~ \isa{ps} ~ \isa{I} ~\isa{r} ~ ((\isa{x}, \isa{r}') \cons \isa{rm}) \,\longleftrightarrow\, \\ 
  & \qquad (\isalet (\isa{resp}, \isa{ps}') = \isa{prv} ~ \isa{I} ~ \isa{x} ~ (\isaco{map} ~ \fst ~ \isa{rm}) ~ \isa{r} ~ \isa{ps} \isain  \\
  & \qquad\:\:  \isalet (ok, I', vs') = \isa{ver}_1 ~ \isa{I} ~ \isa{resp} ~ \isa{r}' ~ \isa{x} ~ (\isaco{map} ~ \fst ~ \isa{rm}) ~ \isa{vs} \isain  \\
  & \qquad\:\:\quad  \isa{ok} \,\land\, \prove ~ \isa{vs}' ~ \isa{prv} ~\isa{ps}' ~ \isa{I}' ~ \isa{r}' ~ \isa{rm})\\
  & \isaend 
\end{align*}
\vspace{-1em}
\caption{Public-coin proofs: protocol definition}
\label{fig:public-coin-proofs-protocol}
\end{figure}

We formalize public-coin proofs in two \emph{locales} in Isabelle/HOL (\Cref{fig:public-coin-proofs-protocol,fig:public-coin-proofs-security}). A locale introduces a context with some variables (\isafixes section) and assumptions about them (\isaassumes section), which can be freely used by definitions and lemmas within that context. Locales can later be instantiated with concrete constants, whereby the assumptions must be proven. In \Cref{sec:generalized-sumcheck}, we will instantiate these locales with the generalized sumcheck protocol and its security proofs.

\subsection{Generic Protocol Definition}
\label{subsec:public-coin-proofs-def}

The first locale, $\isa{public\_coin\_proof}$ (\Cref{fig:public-coin-proofs-protocol}), formalizes the recursive function $\prove$, which models a public-coin proof. The parameters of this function are as follows: 
\begin{itemize}
\item $\isa{ver}_0$ and $\isa{ver}_1$, which are locale parameters, are the verifier functions for the base case and the recursive case,
\item $\isa{vs}$ is the verifier's current state,
\item $\isa{prv}$ and $\isa{ps}$ are the prover and its current state,
\item $\isa{I}$ is the problem instance,
\item $\isa{r}$ is the verifier's randomness for the current round.
\item $\isa{rm}$ is a list of pairs of per-round public information and randomness, which is used for the remaining rounds.
\end{itemize}
The verifier is a locale parameter, as it is part of a protocol definition and remains fixed, while the soundness theorem is a statement about \emph{arbitrary} provers.

The base case, where the list $\isa{rm}$ is empty, determines the verifier's final verdict by calling the verifier function $\isa{ver}_0$. In the recursive case, we first call the prover $\isa{prv}$ and then pass on its response $\isa{resp}$ to the verifier. Here, $(\isaco{map} ~ \fst ~ \isa{rm})$ projects all pairs in $\isa{rm}$ to their first component (i.e., the public information), $\isa{ps}'$ and $\isa{vs}'$ are the prover's and the verifier's successor states, $I'$ is the reduced problem instance, and $\isa{ok}$ is the prover's verdict for the current round. The latter is conjoined to the result of the recursive call to $\prove$ on the prover's and verifier's successor states $\isa{ps}'$ and $\isa{vs}'$, the instance $\isa{I}'$, the randomness $\isa{r}'$ for the next round, and the tail $\isa{rm}$ of the pair list.

There are some differences with \Cref{def:classic_stateful_interaction,def:IPsystems,def:public-coin-proofs}.
First, our prover and verifier functions have additional arguments: the current instance~\isa{I}, the remaining randomness (\isa{r} and second components of $\isa{rm}$), and some additional round-wise public information (first components of $\isa{rm}$). We assume that the initial round-wise public information is derivable from the initial instance~$\isa{I}$. These elements can be easily encoded into the prover and verifier states of \Cref{def:classic_stateful_interaction}.  Second, our prover always gets a random value, even in the first round. This uniformity helps with the inductive security proofs. All these differences are superficial and only introduced for convenience.

\begin{figure}[t]
\begin{align*}
  & \isalocale \isa{public\_coin\_proof\_security} = \\
  & \quad \isa{public\_coin\_proof} ~ \isa{ver}_0 ~ \isa{ver}_1 + {} \\ 
  & \isafixes \isa{S} :: \isa{'i} ~ \isa{set} \\
  & \isaand \isa{honest\_pr} :: (\isa{'i}, \isa{'r}, \isa{'a}, \isa{'resp}, \isa{'ps}) ~ \isa{prv} \\
  & \isaand \isa{sound\_assm} :: \isa{'vs} \fun \isa{'ps} \fun \isa{'i} \fun \isa{'a} ~ \isa{list} \fun \isa{bool} \\
  & \isaand \isa{sound\_err} :: \isa{'i} \fun \isa{real} \\
  & \isaand \isa{compl\_assm} :: \isa{'vs} \fun \isa{'ps} \fun \isa{'i} \fun \isa{'a} ~ \isa{list} \fun \isa{bool} \\
  & \isaand \isa{compl\_err} :: \isa{'i} \fun \isa{real} \\
  & \isaassumes \isa{soundness}: \\
  & \qquad \llbracket \: \isa{I} \notin \isa{S};\; \isa{sound\_assm} ~ \isa{vs} ~ \isa{ps} ~ \isa{I} ~ \isa{xs} \rrbracket \Longrightarrow \\
  & \qquad\quad \Prob_{rs \:\in\: \isaco{tuples}\:(\isaco{length}\:\isa{xs})}[\: \\
  & \qquad\qquad \prove ~\isa{vs} ~ \isa{prv} ~ \isa{ps} ~ \isa{I} ~ \isa{r} ~ (\isaco{zip} ~ \isa{xs} ~ \isa{rs})  \\
  & \qquad\quad \:] \leq \isa{sound\_err} ~ \isa{I} \\
  & \isaand \isa{completeness}:  \\
  & \qquad \llbracket \: \isa{I} \in \isa{S};\; \isa{compl\_assm} ~ \isa{vs} ~ \isa{ps} ~ \isa{I} ~ \isa{xs} \rrbracket \Longrightarrow \\
  & \qquad\quad \Prob_{rs\:\in\: \isaco{tuples}\:(\isaco{length}\:\isa{xs})}[\: \\
  & \qquad\qquad \prove ~\isa{vs} ~ \isa{honest\_pr} ~ \isa{ps} ~ \isa{I} ~ \isa{r} ~ (\isaco{zip} ~ \isa{xs} ~ \isa{rs}) \\
  & \qquad\quad \:] \geq 1 - \isa{compl\_err} ~ \isa{I}  
\end{align*}
\vspace{-.5em}
\caption{Public-coin proofs: security properties}
\label{fig:public-coin-proofs-security}
\end{figure}

\subsection{Generic Security Properties}
\label{subsec:public-coin-proofs-security}

The second locale, $\isa{public\_coin\_proof\_security}$ (\Cref{fig:public-coin-proofs-security}), formalizes the desired security properties of public-coin protocols as assumptions, which must be proven when the locale is instantiated with a concrete protocol. It extends the first locale (indicated by $+$) with additional parameters and with assumptions. The parameters are as follows: $\isa{S}$ is the problem specification (i.e., a set of instances), $\isa{honest\_pr}$ is the honest prover following the protocol, $\isa{sound\_assm}$ and $\isa{compl\_assm}$ are the conditions under which soundness and completeness hold, and $\isa{sound\_err}$ and $\isa{compl\_err}$ are the soundness and completeness errors. 
Finally, the assumptions $\isa{soundness}$ and $\isa{completeness}$ formalize the protocol's security properties as probabilistic statements over the randomness tuples $\isa{rs}$ (cf.~\Cref{def:IPsystems}). Note that, while the initial value~$\SPRandomness$ passed to $\isa{prove}$ need not be random, the successive ones are. One can think of~$\SPRandomness$ as either as an initialization message for the prover or as a dummy value.

We have also defined a variant of this locale, where the completeness is deterministic. We then proved, by instantiating the locale above with the variant, that this corresponds to the special case of the above completeness property where the error is $0$. 

\section{The Generalized Sumcheck Protocol}
\label{sec:generalized-sumcheck}

In this section, we present our formalization of the sumcheck protocol in Isabelle/HOL. We proceed in two stages. First, we axiomatize the required mathematical properties of multivariate polynomials as locale in Isabelle/HOL (\cref{subsec:Assumptions}). 
Then we formalize a generalized sumcheck protocol in the context of that locale (\cref{subsec:protocol-formalization}). Later, we will generically prove its completeness and soundness (\cref{sec:security-properties}) and instantiate our locale with a concrete representation of multivariate polynomials (\cref{sec:polynomial-instantiation}). 
This approach has the advantage of making the proofs more modular and of facilitating future proofs about sumcheck variants over other mathematical structures with properties similar to multivariate polynomials over finite fields such as polynomials over rings and modules, tensor codes, or structures in which the required properties only hold under a cryptographic assumption.

\subsection{Axiomatizing Polynomials} 
\label{subsec:Assumptions}

\begin{figure}[t]
\begin{align*}
  & \isalocale \isa{multi\_variate\_polynomial} = \\
  & \isafixes \\
  & \quad \vars :: (\isa{'p} :: \isa{comm\_monoid\_add}) \fun \isa{'v set} \\
  & \quad \deg :: \isa{'p} \fun \isa{nat} \\
  & \quad \eval :: \isa{'p} \fun (\isa{'v}, \isa{'a} :: \isa{finite})~\isa{subst} \fun \\
  & \qquad\qquad (\isa{'b} :: \isa{comm\_monoid\_add}) \\
  & \quad \inst :: \isa{'p} \fun (\isa{'v}, \isa{'a})~\isa{subst} \fun \isa{'p} \\
  & \isaassumes \\
  & \quad \isa{vars\_finite}: \isaco{finite} ~ (\vars ~ \isa{p}) \\
  & \quad \isa{vars\_zero}: \vars ~ 0 = \emptyset   \\
  & \quad \isa{vars\_add}: \vars ~ (\isa{p} + \isa{q}) \subseteq \vars ~ \isa{p}  \cup \vars ~ \isa{q}  \\
  & \quad \isa{vars\_inst}: \vars ~ (\inst ~ \isa{p} ~ \mysigma) \subseteq \vars ~ \isa{p} \setminus \dom ~ \mysigma \\[.6ex]
  & \quad \isa{deg\_zero}: \deg ~ 0 = 0  \\
  & \quad \isa{deg\_add}: \deg ~ (\isa{p} + \isa{q}) \leq \isaco{max} ~ (\deg ~ \isa{p}) ~ (\deg ~ \isa{q}) \\
  & \quad \isa{deg\_inst}: \deg ~ (\inst ~ \isa{p} ~ \mysigma) \leq \deg ~ \isa{p} \\[.6ex]
  & \quad \isa{eval\_zero}: \eval ~ 0 ~ \mysigma = 0 \\
  & \quad \isa{eval\_add}: \vars ~ \isa{p} \cup \vars ~ \isa{q} \subseteq \dom ~ \mysigma  \\
  & \qquad\quad\;\Longrightarrow 
    \eval ~ (\isa{p} + \isa{q}) ~ \mysigma = \eval ~ \isa{p} ~ \mysigma + \eval ~ \isa{q} ~ \mysigma  \\
  & \quad \isa{eval\_inst}: \vars ~ \isa{p} \subseteq \dom ~ \mysigma \cup \dom ~ \myrho  \\
  & \qquad\quad\;\Longrightarrow 
    \eval ~ (\inst ~ \isa{p} ~ \mysigma) ~ \myrho = \eval ~ \isa{p} ~ (\myrho \mapupdate \mysigma) \\[.6ex]
  & \quad \isa{roots}: \lvert\: \{r \mid 
       \vars ~ \isa{p} \subseteq \{x\} \:\land\: \vars ~ \isa{q} \subseteq \{x\}  \\
  & \phantom{\quad \isa{roots}: \lvert\: \{ } 
       \:\land\: \deg ~ \isa{q} \leq \isa{d} \:\land\:  \deg ~ \isa{p} \leq \isa{d}  
       \:\land\: \isa{p} \neq \isa {q} \:\land\: \\
  & \phantom{\quad \isa{roots}: \lvert\: \{ } 
    \:\land\: \eval ~ \isa{p} ~ [\isa{x} \mapsto \SPRandomness] 
            = \eval ~ \isa{q} ~ [\isa{x} \mapsto \SPRandomness] \} \:\rvert \leq \isa{d} 
\end{align*}
\vspace{-.5em}
\caption{The locale abstracting multivariate polynomials}
\label{fig:locale}
\end{figure}

We formalize our assumptions about multivariate polynomials in an Isabelle/HOL locale (\Cref{fig:locale}). This locale declares four function variables along with their types: given a polynomial $\isa{p}$ of abstract type \isa{'p} and a substitution (i.e., partial function) $\mysigma$ from variables of type \isa{'v} to values of type \isa{'a}, 
\begin{enumerate}
\item $\vars~\isa{p}$ denotes the set of $\isa{p}$'s variables, 
\item $\deg~\isa{p}$ is $\isa{p}$'s degree, 
\item $\eval~\isa{p}~\mysigma$ evaluates $\isa{p}$ to a value of type \isa{'b} using $\mysigma$, and 
\item $\inst~\isa{p}~\mysigma$ instantiates $\isa{p}$ to another polynomial using $\mysigma$.
\end{enumerate}

In contrast to the common formulation of the sumcheck protocol over a finite field~$\F$, we (only) impose an (additive) commutative monoid structure on the types of polynomials~\isa{'p} and their values~\isa{'b}, and we assume that the type~\isa{'a} of their arguments is finite. In Isabelle, we can achieve this by adding type class constraints to these type variables (e.g., $\isa{'p} :: \isa{comm\_monoid\_add}$).
Note that $\eval$ has different argument and result types $\isa{'a}$ and $\isa{'b}$. While these types coincide for polynomials over a field $\F$ as used in \Cref{prot:recursive_protocol}, they may differ for other instantiations. For example, one could define a sumcheck protocol for polynomials over \emph{modules} over a ring, in which case the arguments would be ring elements and the results would be module elements.

Each of the four function variables comes with a set of assumptions about it. Most of these concern the interaction of these functions with the additive structure and with instantiation. Here, we only discuss the final three in more detail.
The assumption \isa{eval\_add} states that the result of evaluating the sum $\isa{p}+\isa{q}$ of two polynomials under a substitution $\mysigma$ whose domain contains all their variables is the same as adding the results of their individual evaluation under $\mysigma$. 
The assumption \isa{eval\_inst} relates evaluation and instantiation and expresses that, given two substitutions $\mysigma$ and $\myrho$ whose domains jointly cover $\isa{p}$'s variables, first instantiating a polynomial~$\isa{p}$ with $\mysigma$, followed by an evaluation under $\myrho$ is the same as evaluating $\isa{p}$ under the substitution $\myrho \mapupdate \mysigma$, which updates the substitution $\myrho$ with $\mysigma$.
The final assumption, \isa{roots}, states that any pair of different univariate polynomials $\isa{p}$ and $\isa{q}$ of degrees bounded by $\isa{d}$ are equal on at most $\isa{d}$ points. We could equivalently assume that the number of roots of a univariate polynomial is bounded by its degree. This constitutes the main assumption required for the soundness proof.

Apart from the standard instance of multivariate polynomials over finite fields, discussed in \Cref{sec:polynomial-instantiation}, multivariate polynomials over certain rings satisfy small variants of these assumptions, where the roots assumption is modified so that polynomials are only required to have a bounded number of roots in a special \emph{sampling set}~\cite{CCKP19}, which is a subset of the ring. The verifier's random messages in the sumcheck protocol must then be drawn from the sampling set. Similar considerations hold for modules \cite{scarguments}. Therefore, we expect that it would take little effort to extend our formalization to work with sumcheck variants using polynomials over rings and modules with sampling sets.

We proved a collection of lemmas derived from the locale's assumptions. These include finite sum variants of the assumptions involving binary addition and three lemmas about the interaction of evaluation, instantiation, and finite sums, which we present below.

The first lemma, \isa{eval\_sum\_inst}, simplifies the evaluation of a finite sum of instantiations to a finite sum of evaluations.
\begin{align*}
& \isalemma \isa{eval\_sum\_inst}: \\ 
& \quad \isaassumes \vars ~ \isa{p} \subseteq \isa{V} \,\cup\, \dom ~ \myrho  \isaand  \isaco{finite} ~ \isa{V} \\
& \quad \isashows \eval ~ (\sum_{\sigma \,\in\, \substs \: \isa{V} \: \isa{H}} \inst ~ \isa{p} ~ \mysigma) ~ \myrho \\
& \hspace{13mm}
                = (\sum_{\sigma \,\in\, \substs \: \isa{V} \: \isa{H}} \eval ~ \isa{p} ~ (\myrho \mapupdate \mysigma)) 
\end{align*}

The second lemma, $\isa{eval\_sum\_inst\_commute}$, simultaneously commutes evaluation with sum and swaps the substitutions used in the evaluation and the instantiation, respectively.
\begin{align*}
& \isalemma \isa{eval\_sum\_inst\_commute}: \\ 
& \quad \isaassumes  \vars ~ \isa{p} \subseteq \{ \isa{x} \} \cup \isa{V}  \isaand \isa{x} \notin \isa{V} \isaand \isaco{finite} ~ \isa{V} \\ 
& \quad \isashows \eval~ ( \sum_{\sigma \,\in\, \isa{substs} \: \isa{V} \: \isa{H}} \inst ~ p ~ \mysigma ) ~  [\isa{x} \mapsto \isa{r}] \\
& \phantom{\quad \isashows }
  =  \sum_{\sigma \,\in\, \isa{substs} \: \isa{V} \: \isa{H}} \eval ~ (\inst ~ \isa{p} ~ [\isa{x} \mapsto {r}] ) ~ \mysigma  
\end{align*}

The third lemma, \isa{sum\_merge}, combines two sums into one, merging a singleton substitution into one with an extended domain. 
\begin{align*}
& \isalemma \isa{sum\_merge}: \\ 
& \quad \isaassumes  \isa{x} \notin \isa{V} \\ 
& \quad \isashows \sum_{\isa{h} \,\in\, \isa{H}} (\sum_{\sigma \in \isa{substs} \: \isa{V} \: \isa{H} } \eval ~ \isa{p} ~ ([\isa{x} \mapsto \isa{h}] ~ \texttt{++} ~ \mysigma )) \\
& \phantom{\quad \isashows}
  = \sum_{\sigma \,\in\, \isa{substs} \: (\{\isa{x}\} \cup \isa{V}) \: \isa{H}} \eval ~ \isa{p} ~ \mysigma  
\end{align*}

The proofs of the above three lemmas use basic facts about sums and maps together with our assumptions on $\vars$, $\eval$, and $\inst$ and their aforementioned extensions to finite sums.

\subsection{Protocol Formalization}
\label{subsec:protocol-formalization}

Since we have fixed the types of arguments~\isa{'a} and results~\isa{'b} of our abstract polynomials~\isa{'p} in our locale, we will henceforth describe sumcheck instances by triples $(\SPSubset, \SPPolynomial, \SPTarget) \in \SPLanguage$ (omitting the finite field $\F$). In Isabelle/HOL, we define $\SPLanguage$ as follows.
\begin{align*}
  & \isatypesynonym (\isa{'p}, \isa{'a}, \isa{'v}) ~ \isa{sc\_inst} = \isa{'a set} \times \isa{'p} \times \isa{'v} \\[1ex]
  & \isadefinition \SPLanguage :: (\isa{'p}, \isa{'a}, \isa{'v}) ~ \isa{sc\_inst} ~ \isa{set} \isawhere \\
  & \quad \SPLanguage = \{(\isa{H}, \isa{p}, \isa{v}) \mid 
          \sum_{\sigma \,\in\, \substs \: (\vars \: p) \: H} \eval ~ p ~\mysigma = \isa{v} \}
\end{align*}

For better readability and easier reference, we formalize the generalized sumcheck protocol in Isabelle/HOL as a separate recursively defined predicate $\sumcheck$, which invokes the prover and integrates the verifier. 
However, we do also define the sumcheck protocol as an instance of our generic definition of public-coin proofs (cf.~\Cref{fig:public-coin-proofs-protocol}) and establish its equality with the definition given here. 

The predicate $\sumcheck$ is parameterized by the prover, which is a function of type 
\begin{align*}
    & \isatypesynonym (\isa{'p}, \isa{'a}, \isa{'b}, \isa{'v}, \isa{'s}) ~ \isa{prover} = \\
    & \quad (\isa{'p}, \isa{'a}, \isa{'b}) ~ \isa{sc\_inst} \fun \isa{'v} \fun \isa{'v list} \fun \isa{'a} \fun \isa{'s} \fun \isa{'p} \times \isa{'s}
\end{align*} 
Note that this type specializes the prover type $\isa{prv}$ from \Cref{fig:public-coin-proofs-protocol}.
For a prover $\SPProver$ of this type, the application 
$
(\SPProver ~ \isa{I} ~ \isa{x} ~ \isa{xs} ~ \SPRandomness ~ \SPState)
$ 
returns a pair $(\hat{\SPMessagePolynomial}, \NewSPState)$ consisting of a univariate polynomial~$\hat{\SPMessagePolynomial}$, which is sent as the next challenge to the verifier, and the prover's successor state $\NewSPState$. The parameters are as follows:  $\isa{I}$ is the sumcheck instance, $\isa{x}$ is the returned polynomial $\hat{\SPMessagePolynomial}$'s expected variable, $\isa{xs}$ is a list of variables not containing $\isa{x}$ such that $\vars ~ \SPPolynomial \subseteq \setfun ~ \isa{xs} \,\cup \{\isa{x}\}$, $\SPRandomness$ is the verifier's next random value, and $\SPState$ is the prover's current state.

\begin{figure}
\begin{align*}
    \isafun &\sumcheck :: (\isa{'v}, \isa{'a}, \isa{'p}, \isa{'s})~\isa{prover} \fun \isa{'s} \fun \isa{'a set} \fun \isa{'p} \fun \\
    & \hspace{18mm}\isa{'b} \fun \isa{'a} \fun (\isa{'v} \times \isa{'a})~\isa{list} \fun \isa{bool} \isawhere \\
    &\sumcheck ~ \SPProver ~ \SPState ~ (\SPSubset, \SPPolynomial, \SPTarget) ~ \SPRandomness ~ [] \,\longleftrightarrow\, 
    \SPTarget = \eval ~ \SPPolynomial ~ \mapempty \\
    \mid\; &\sumcheck ~ \SPProver ~ \SPState ~ (\SPSubset, \SPPolynomial, \SPTarget) ~ \SPRandomness ~ ((\isa{x}, \NewSPRandomness) \cons \isa{rm}) \,\longleftrightarrow\, \\
    &\quad \isalet (\hat{\SPMessagePolynomial}, \NewSPState) = \SPProver ~ (\SPSubset, \SPPolynomial, \SPTarget) ~ \isa{x} ~ (\mapfun ~ \fst ~ \isa{rm}) ~ \SPRandomness ~ \SPState \isain \\
    &\qquad\vars ~ \hat{\SPMessagePolynomial} \subseteq \{\isa{x}\} \isand \deg ~ \hat{\SPMessagePolynomial} \leq \deg ~ \SPPolynomial \:\land {} \\
    &\qquad \SPTarget = \sum_{\isa{h} \in \SPSubset} \eval ~ \hat{\SPMessagePolynomial} ~ [\isa{x} \mapsto \isa{h}] \:\land {}\\
    &\qquad \sumcheck ~ \SPProver ~ \NewSPState ~ \\
    &\qquad\quad (\SPSubset,\, 
                   \inst ~ \SPPolynomial ~ [\isa{x} \mapsto \NewSPRandomness],\, 
                   \eval ~ \hat{\SPMessagePolynomial} ~ [\isa{x} \mapsto \NewSPRandomness]) ~ \NewSPRandomness ~ \isa{rm} 
\end{align*}
    \caption{Formal definition of the sumcheck protocol}
    \label{fig:sumcheck-def}
\end{figure}

We then define the predicate $\sumcheck$ as in \Cref{fig:sumcheck-def}.
An application $(\sumcheck ~ \SPProver ~ \SPState ~ (\SPSubset,\; \SPPolynomial,\; \SPTarget) ~ \SPRandomness ~ \isa{rm})$ returns the verifier's verdict (accept or reject). The parameters are as follows: $\SPProver$ is the prover as described above, 
$\SPState$ is the prover's current state, $(\SPSubset, \SPPolynomial, \SPTarget)$ describe the current sumcheck instance, $\SPRandomness$ is the randomness sent by the verifier as an input to the prover, and $\isa{rm}$ is an ordered substitution consisting of a list of pairs of variables and random values. 

The predicate $\sumcheck$ is defined using two equations (i.e., logical equivalences). The first equation describes the base case where the list $\isa{rm}$ is empty and the polynomial $\SPPolynomial$ is a constant, which is evaluated under the empty substitution $\mapempty$ and compared to the value $\SPTarget$.
The second equation describes the recursive case, which reduces the instance by executing one round of the protocol. 
We first call the prover $\SPProver$ to obtain the prover's message to the verifier, i.e., the polynomial $\hat{\SPMessagePolynomial}$, and its successor state $\NewSPState$. 
We call the projection $(\mapfun ~ \fst ~ \isa{rm})$ of $\isa{rm}$ to the public information the domain of $\isa{rm}$. 
The verifier then performs the following checks: 
\begin{enumerate}[label=(\roman*)]
\item \emph{variable check}: the received polynomial $\hat{\SPMessagePolynomial}$ has (at most) the expected variable, 
\item \emph{degree check}: $\hat{\SPMessagePolynomial}$'s degree is bounded by $\SPPolynomial$'s degree, and 
\item \emph{evaluation check}: the sum of $\hat{\SPMessagePolynomial}$ over all $\isa{h} \in \SPSubset$ matches the value $\SPTarget$.
\end{enumerate}
These checks are conjoined with a recursive call of the function $\sumcheck$ with the new state $\NewSPState$ of the prover and the reduced sumcheck instance $(\SPSubset, \NewSPPolynomial, \NewSPTarget)$, where the polynomial $\NewSPPolynomial = \inst ~ \SPPolynomial ~ [\isa{x} \mapsto \NewSPRandomness]$ instantiates $\SPPolynomial$'s variable $\isa{x}$ to $\NewSPRandomness$ and the value $\NewSPTarget = \eval ~ \hat{\SPMessagePolynomial} ~ [\isa{x} \mapsto \NewSPRandomness]$ evaluates $\SPMessagePolynomial$ at $\NewSPRandomness$. We call this the \emph{recursive check}. 
Note that the recursion terminates, since the length of the list $\isa{rm}$ decreases in each call.

Finally, we define the honest prover to match the protocol description (cf.~\cref{prot:recursive_protocol}): 
\begin{align*}
    & \isadefinition \honestprover :: (\isa{'v}, \isa{'a}, \isa{'p}, \isa{unit})~\isa{prover} \isawhere \\
    & \quad \honestprover ~ (\SPSubset,\; \SPPolynomial,\; \isa{v}) ~ \isa{x} ~ \isa{xs} ~ \isa{r} ~ \_ = \\
    & \qquad (\sum_{\sigma \,\in\, \substs \: (\setfun \: \isa{xs}) \: \SPSubset} \inst ~ \SPPolynomial ~ \mysigma, ())
\end{align*}
This function returns the univariate polynomial $\SPMessagePolynomial = (\sum_{\sigma \,\in\, \substs \: (\setfun \: \isa{xs}) \: \SPSubset} \inst ~ \SPPolynomial ~ \mysigma)$ with variable $\isa{x} \notin \setfun ~ \isa{xs}$, obtained by summing up all instances of $\SPPolynomial$ over all substitutions $\mysigma$ assigning values in $\SPSubset$ to the variables in $\setfun ~ \isa{xs}$. This prover does not use a state (hence the state's type is $\isa{unit}$) and ignores the sumcheck instance's value $\isa{v}$, the current variable $\isa{x}$, and the randomness $\isa{r}$ sent by the verifier.

\subsection{Discussion of our Formalization}

Comparing with common informal accounts of the sumcheck protocol (see, e.g., \cite{AroraBarak2009,Thaler2022}), we find the following additional points noteworthy. 
First, we formalize the entire interaction between the prover and the verifier as a single recursive predicate $\isaco{sumcheck}$, which is actually an instance of a general definition (cf.~\Cref{fig:public-coin-proofs-protocol}), thus replacing an informal sequence of protocol steps with a compact and rigorous mathematical description.
Second, we model randomness as externally provided rather than generated on-the-fly, which allows for a simpler formalization and proofs based on deterministic functions rather than probabilistic monads.
Third, we model a generalized sumcheck protocol, where we abstract the underlying structure of polynomials by a set of functions and assumptions over them (\Cref{fig:locale}).

We expect that our formalization would extend easily to various generalizations of the language $\SPLanguage$ and the sumcheck protocol. This includes instances where the polynomial $\SPPolynomial(\XVar_1,\ldots,\XVar_\SPArity)$ is evaluated and summed over possibly different sets $\SPSubset_1,\ldots,\SPSubset_\SPArity$ for each variable $\XVar_1,\ldots,\XVar_\SPArity$, to check whether
\begin{equation*}
    \SPTarget = \sum_{\SumVal_1\in\SPSubset_1,\ldots,\SumVal_\SPArity \in \SPSubset_\SPArity} \SPPolynomial(\SumVal_1, \SumVal_2, \ldots, \SumVal_\SPArity)
    \enspace.
\end{equation*}
One could also consider weighted sums
\begin{equation*}
    \SPTarget = \sum_{(\SumVal_1,\ldots,\SumVal_\SPArity) \in \SPSubset^\SPArity} w_1(\SumVal_1)\cdots w_\SPArity(\SumVal_\SPArity) \cdot\SPPolynomial(\SumVal_1, \SumVal_2, \ldots, \SumVal_\SPArity)
\end{equation*}
for weight functions $w_1,\ldots,w_\SPArity\colon\SPSubset\to\F$ as in \cite{tensor1}, by including $w_2,\ldots,w_\SPArity$ in the sum defining $\SPMessagePolynomial$ and changing the verifier checks to  $\SPTarget \CheckEq \sum_{\SumVal\in \SPSubset} w_1(\SumVal_1)\SPMessagePolynomial(\SumVal)$. Setting the weight functions appropriately allows the sumcheck protocol to work with multisets.

\section{Security Properties}
\label{sec:security-properties}

We now turn to the two main results that we proved about the sumcheck protocol: its completeness and soundness. We have also instantiated the locale $\isa{public\_coin\_proof\_security}$ from \Cref{fig:public-coin-proofs-security} with our security results, thus showing that they match the general definitions. 
We start our presentation with the (simpler) completeness property.

\subsection{Completeness}
\label{subsec:completeness}

The sumcheck protocol's completeness theorem (cf.~\Cref{thm:paper-sumcheck-theorem}) states that the verifier will (deterministically) accept any run with the honest prover on an instance $(\SPSubset, \SPPolynomial, \SPTarget) \in \SPLanguage$ of the sumcheck problem. We formalize this theorem as follows. 
\begin{align*}
  & \isatheorem \isa{completeness}:\\
  & \quad\isaassumes \\ 
  & \qquad (\SPSubset, \SPPolynomial, \SPTarget) \in \SPLanguage \\
  & \qquad \vars ~ \SPPolynomial = \setfun ~ (\mapfun ~ \fst ~ \isa{rm}) \\
  & \qquad \distinct ~ (\mapfun ~ \fst ~ \isa{rm}) \\
  & \qquad \SPSubset \neq \emptyset \\
  & \quad \isashows \sumcheck ~ \honestprover ~ \SPState ~ (\SPSubset, \SPPolynomial, \SPTarget) ~ \SPRandomness ~ \isa{rm} 
\end{align*}
The additional assumptions state that (i) the set of $\SPPolynomial$'s variables corresponds to the domain of $\isa{rm}$, (ii) the variables in the domain of $\isa{rm}$ are all distinct, and (iii) that the set $\SPSubset$ is non-empty. 

For the inductive proof to succeed, we slightly generalize this statement as follows.
\begin{align*}
  & \isalemma \isa{completeness\_inductive}:\\
  & \quad\isaassumes \\ 
  & \qquad \SPTarget \,= \sum_{\sigma \,\in\, \substs \: (\setfun \: (\mapfun \: \fst \: \isa{rm}))\:\SPSubset} \eval ~ \SPPolynomial ~ \mysigma \\
  & \qquad \vars ~ \SPPolynomial \subseteq \setfun ~ (\mapfun ~ \fst ~ \isa{rm}) \\
  & \qquad \distinct ~ (\mapfun ~ \fst ~ \isa{rm}) \\
  & \qquad \SPSubset \neq \emptyset \\
  & \quad \isashows \\
  & \qquad \sumcheck ~ \honestprover ~ \SPState ~ (\SPSubset, \SPPolynomial, \SPTarget) ~ \SPRandomness ~ \isa{rm} 
\end{align*}
Here, the equality in the second assumption has turned into a set inclusion. The reason is that the arity of $\SPPolynomial$ may decrease by more than one in each recursive call, while the length of $\isa{rm}$ decreases by exactly one. For example, $p=x y$ has arity $2$, while $(\isa{inst}~p~[x \mapsto 0])=0$ has arity $0$.
As a result, we have also unfolded the definition of $\SPLanguage$ and replaced the domain $\vars ~ p$ of the substitutions in the sum by the overapproximation $\setfun ~ (\mapfun ~ \fst ~ \isa{rm})$. We now prove this lemma.

\begin{proof} 
Our proof proceeds by computation induction on the definition of $\sumcheck$ (\cref{subsec:protocol-formalization}).\footnote{An induction on $\isa{rm}$ would also work, but would be more tedious.} This form of induction exactly follows the structure of the function definition, for which, crucially, termination is proven as part of its definition. Isabelle automatically generates a computation induction rule for each recursively defined function. 

\medskip

\emph{Base case.} For this case, we must show that 
$$
\sumcheck ~ \honestprover ~ \SPState ~ (\SPSubset, \SPPolynomial, \SPTarget) ~ \SPRandomness ~ []
$$ 
holds, which Isabelle's simplifier proves automatically. Since $\isa{rm} = []$, the goal simplifies to $\SPTarget = \eval ~ \SPPolynomial ~ \mapempty$, as does the theorem's first assumption, since $\substs ~ \emptyset ~ \SPSubset = \{\mapempty\}$. 

\medskip

\emph{Inductive step.} In the inductive step, we have to show that
\begin{equation} \label{eq:compl-ind-step}
\sumcheck ~ \honestprover ~ \SPState ~ (\SPSubset, \SPPolynomial, \SPTarget) ~ \SPRandomness ~ ((\isa{x}, \NewSPRandomness) \cons \isa{rm}')
\end{equation} 
holds under the assumptions 
\begin{align}
  \label{ass:comp1} 
  & \SPTarget = \sum_{\sigma \,\in\, \substs\:(\setfun\:(\mapfun\:\fst\:((\isa{x}, \NewSPRandomness) \cons \isa{rm'})))\:\SPSubset} \eval ~ \SPPolynomial ~ \mysigma \\
  \label{ass:comp2} 
  & \vars ~ \SPPolynomial \subseteq \setfun ~ (\mapfun ~ \fst ~ ((\isa{x}, \NewSPRandomness) \cons \isa{rm}')) \\
  \label{ass:comp3} 
  & \distinct ~ (\mapfun ~ \fst ~ ((\isa{x}, \NewSPRandomness) \cons \isa{rm}')) \\
  \label{ass:comp4} 
  & \SPSubset \neq \emptyset
\end{align}
and the induction hypothesis (see below).
Let $V = \setfun ~ (\mapfun ~ \fst ~ \isa{rm}')$.
By definition of the $\sumcheck$ function, our goal \eqref{eq:compl-ind-step} reduces to proving that the variable, degree, evaluation and recursive checks succeed given 
\begin{align*}
(\hat{\SPMessagePolynomial}, \NewSPState) 
& = \honestprover ~ \SPSubset ~ \SPPolynomial ~ \isa{x} ~ V ~ \SPRandomness ~ \SPState \\
& = (\sum_{\sigma \,\in\, \substs \: V \: \SPSubset} \inst ~ \SPPolynomial ~ \mysigma, ()).
\end{align*}

The variable and degree checks, $\vars ~ \hat{\SPMessagePolynomial} \subseteq \{\isa{x}\}$ and $\deg ~ \hat{\SPMessagePolynomial} \leq \deg ~ \SPPolynomial $, are straightforward using the definition of the honest prover, locale assumptions on $\vars$, $\deg$, and $\inst$ and the assumptions \eqref{ass:comp2} and \eqref{ass:comp4} above.

The evaluation check, namely, that $\SPTarget = \sum_{\isa{h} \in \SPSubset} \eval ~ \hat{\SPMessagePolynomial} ~ [\isa{x} \mapsto \isa{h}]$, follows the chain of equations
\begin{align}
\label{eq:comp0}
\SPTarget &= \sum_{\sigma \,\in\, \substs \: (V \: \cup \: \{\isa{x}\}) \: \SPSubset} \eval ~ \SPPolynomial ~ \mysigma \\
\label{eq:comp1}
&= \sum_{\isa{h} \,\in\, \SPSubset} \left(\sum_{\sigma \,\in\, \substs \: V \: \SPSubset} \isa{eval p} \: ([\isa{x} \mapsto \isa{h}] \mapupdate \mysigma) \right)\\
\label{eq:comp2}
&= \sum_{\isa{h} \,\in\, \SPSubset} \eval ~ \left(\sum_{\sigma \,\in\, \substs \: V \: H} \inst ~ \SPPolynomial ~ \mysigma \right) ~ [\isa{x} \mapsto \isa{h}] \\
\label{eq:comp3}
&= \sum_{\isa{h} \,\in\, \SPSubset} \eval ~ \hat{\SPMessagePolynomial} ~ [\isa{x} \mapsto \isa{h}],
\end{align}
where 
\Cref{eq:comp0} comes from assumption \eqref{ass:comp1}, 
\Cref{eq:comp1} uses $\isa{sum\_merge}$ (\cref{subsec:Assumptions}) and $\isa{x} \notin \isa{V}$, which follows from assumption \eqref{ass:comp3},
\Cref{eq:comp2} uses the lemma $\isa{eval\_sum\_inst}$ (\cref{subsec:Assumptions}) and assumption \eqref{ass:comp2} and 
\Cref{eq:comp3} holds by definition of $\hat{q}$. 

Finally, the recursive check is 
\begin{align*}
&\sumcheck ~ \honestprover ~ \NewSPState ~ \isa{I}' ~ \NewSPRandomness ~ \isa{rm}',
\end{align*}
where $\isa{I}' = (\SPSubset, \inst ~ \SPPolynomial ~ [\isa{x} \mapsto \NewSPRandomness],  
    \eval ~ \hat{\SPMessagePolynomial} ~ [\isa{x} \mapsto \NewSPRandomness])$.
This is exactly the conclusion of the induction hypothesis, which holds under the assumptions 
\begin{align}
  & \eval ~ \hat{\SPMessagePolynomial} ~ [\isa{x} \mapsto \NewSPRandomness] \: = \sum_{\sigma \,\in\, \substs \: V \: \SPSubset} \eval ~ (\inst ~ \SPPolynomial ~ [\isa{x} \mapsto \NewSPRandomness] ) ~ \mysigma \\
  & \vars ~ (\inst ~ \SPPolynomial ~ [\isa{x} \mapsto \NewSPRandomness]) \subseteq \dom ~ (\mapof ~ \isa{rm}') \\
  & \distinct ~ (\mapfun ~ \fst ~ \isa{rm}') \\
  & \SPSubset \neq \emptyset.
\end{align}
Using the lemma $\isa{eval\_sum\_inst\_commute}$ (\Cref{subsec:Assumptions}), we prove the first assumption and the remaining ones follow straightforwardly from the locale assumptions and assumptions \eqref{ass:comp1}-\eqref{ass:comp4} of the induction step, and therefore so does the recursive check.  
\end{proof}

\subsection{Soundness}
\label{subsec:soundness}

We formalize the soundness theorem of the sumcheck protocol (cf.~\cref{thm:paper-sumcheck-theorem})   as follows.
\begin{align*}
    &\isatheorem \isa{soundness}:\\
    & \quad \isaassumes (\SPSubset, \SPPolynomial, \SPTarget) \notin \SPLanguage \\
    & \qquad \isaand
      \vars ~ \SPPolynomial = \isaco{set} ~ \isa{vs} \isaand 
      \distinct ~ \isa{vs} \isaand
      H \neq \emptyset  \\
    & \quad \isashows 
      \Prob_{\isa{rs} \,\in\, \tuples \: (\isaco{arity} \: \SPPolynomial)}[\: \\
    & \qquad\qquad\quad 
         \sumcheck ~ \SPProver ~ \SPState ~ (\SPSubset, \SPPolynomial, \SPTarget) ~ \SPRandomness ~ (\isaco{zip} ~ \isa{vs} ~ \isa{rs}) \\
    & \qquad\qquad 
       \:] \leq \frac{\deg ~ \SPPolynomial \cdot \isaco{arity} ~ \SPPolynomial}{\card{\UnivA}}
\end{align*}

Some explanatory remarks are in order here. First, the randomness is taken over a tuple $\isa{rs}$ of length the arity of $\SPPolynomial$, where $\isaco{arity}~\SPPolynomial$ is defined as the size of $\vars~\SPPolynomial$. Recall that, in our protocol formalization, the prover always gets an input, so $\SPRandomness$ corresponds to an initial value sent by the verifier. Although the initial value $\SPRandomness$ need not be random, those passed in subsequent recursive calls will be. This structure makes all rounds look the same, which helps when setting up the induction proof. Second, note that the first two assumptions imply that $\isa{vs}$ has the same length as $\isa{rs}$. Hence, all list elements are combined into pairs in $(\isaco{zip} ~ \isa{vs} ~ \isa{rs})$. Third, we use $\UnivA$ to denote the  (finite) set of all elements of the type $\isa{'a}$ and $\card{\UnivA}$ is its cardinality.

Unlike the completeness statement above we cannot prove this theorem by computation induction on the definition of $\sumcheck$, since $\isa{rs}$ in its last argument $(\isaco{zip} ~ \isa{vs}~ \isa{rs})$ is bound by the probability statement. Instead, we use induction on the variable list $\isa{vs}$ and, since the variables $\SPPolynomial$, $\SPTarget$, $\SPState$, and $\SPRandomness$ change in recursive calls to $\sumcheck$, we generalize the induction over these variables, meaning that they appear universally quantified in the induction hypothesis and can thus be instantiated with different values in the proof.

Moreover, similar to the completeness theorem, this theorem's second assumption, $\vars ~ \SPPolynomial = \isaco{set} ~ \isa{vs}$, is too strong, 
and must be weakened to $\vars ~ \SPPolynomial \subseteq \isaco{set} ~ \isa{vs}$. We also replace $\vars ~ \SPPolynomial$ by $\isaco{set} ~ \isa{vs}$ and $\isaco{arity} ~ \SPPolynomial$ by $\isaco{length} ~ \isa{vs}$ in the conclusion. Similarly, the degree of the reduced instance's $\NewSPPolynomial$ might be (possibly much) smaller than $\SPPolynomial$'s degree. Hence, we introduce an additional assumption $\deg ~ \SPPolynomial \leq \SPDegree$ with an upper bound $\SPDegree$ for the degree, which replaces $\deg ~ \SPPolynomial$ in the theorem's soundness error. 

This discussion leads to the following inductive generalization of the soundness theorem for which the sketched induction proof setup succeeds. 
\begin{align*}
    &\isalemma \isa{soundness\_inductive}:\\
    & \quad \isaassumes \\
    & \qquad \deg ~ \SPPolynomial \leq \SPDegree \isaand \vars ~ \SPPolynomial \subseteq \isaco{set} ~ \isa{vs} \isaand {} \\
    & \qquad \distinct ~ \isa{vs} \isaand 
      \SPSubset \neq \emptyset  \\
    & \quad \isashows \\
    & \qquad \Prob_{\isa{rs} \:\in\: \tuples \: (\isaco{length} \: \isa{vs})}[ \\
    & \qquad\quad \sumcheck ~ \SPProver ~ \SPState ~ (\SPSubset, \SPPolynomial, \SPTarget) ~ \SPRandomness ~ (\isaco{zip} ~ \isa{vs} ~ \isa{rs}) \;\land\; \\
    & \qquad\quad 
    v \neq \sum_{\sigma \: \in \: \substs \: (\isaco{set} \: \isa{vs}) \: H} \eval ~ \SPPolynomial ~ \mysigma  \\
    & \qquad ] \leq \frac{\SPDegree \cdot \isaco{length} ~ \isa{vs}}{\card{\UnivA}}
\end{align*}
\begin{proof}
By list induction on $\isa{vs}$, generalizing over $\SPState$, $\SPPolynomial$, $\SPTarget$, and $\SPRandomness$.

\medskip
\emph{Base case.} 
For $\isa{vs} = []$, we have to establish that 
\begin{align*}
    & \Prob_{rs \:\in\: \tuples \; 0}[\: \\
    & \quad\; 
      \sumcheck ~ \SPProver ~ \SPState ~ \SPInstanceTupleIsa ~ \SPRandomness ~ [] \,\land\, \\
    & \quad\;
      \SPTarget \neq \eval ~ \SPPolynomial ~ \emptyset  \\
    & \:] \leq 0
\end{align*}
which is proven automatically by Isabelle's simplifier, since $(\sumcheck ~ \SPProver ~ \SPState ~ (\SPSubset,\; \SPPolynomial,\; \SPTarget) ~ \SPRandomness ~ [])$ reduces to $\SPTarget = \eval ~ \SPPolynomial ~ \emptyset$ by the definition of $\sumcheck$.

\medskip
\emph{Inductive step.} 
For $\isa{vs}=\isa{x} \cons \isa{vs}'$, we have to show that 
\begin{align*}
    & \Prob_{\SPRandomness_1\cons\isa{rs} \:\in\: \tuples \: (\isaco{Suc} \: (\isaco{length} \: \isa{vs}'))}[ \\
    & \quad\; \sumcheck ~ \SPProver ~ \SPState ~ \SPInstanceTupleIsa ~ \SPRandomness ~ (\isaco{zip} ~ (\isa{x} \cons \isa{vs}') ~ (\SPRandomness_1\cons\isa{rs})) \:\land\: {} \\
    & \quad\; 
    \SPTarget \:\neq \sum_{\sigma \: \in \: \substs \: (\{\isa{x}\} \,\cup\, \isaco{set} \: \isa{vs}') \: \SPSubset} \eval~ \SPPolynomial ~ \mysigma  \\
    & ] \leq \frac{\SPDegree \cdot \isaco{Suc} ~ (\isaco{length} ~ \isa{vs}')}{\card{\UnivA}}
\end{align*}
under the assumptions $\vars ~ \SPPolynomial \subseteq \{\isa{x}\} \cup \isaco{set} ~ \isa{vs}'$, $\distinct ~ (\isa{x} \cons \isa{vs}')$ and the remaining two (unchanged) assumptions of the lemma. 

To make the proof more readable, we introduce some abbreviations. We abbreviate the (unfolded) language non-membership condition to 
\begin{align*}
   \failure & \:\longleftrightarrow\:  \SPTarget \:\neq \sum_{\sigma \: \in \: \substs \: (\{\isa{x}\} \,\cup\, \isaco{set} \: \isa{vs}') \: \SPSubset} \eval ~ \SPPolynomial ~ \mysigma 
\end{align*}       
and we also introduce abbreviations the verifier's degree, variable, and evaluation checks and for the recursive check:
\begin{align*}
\verchecks & \:\longleftrightarrow\: \deg ~ \Mal{\isa{q}} \leq  \deg ~ \SPPolynomial \isand \vars ~ \hat{\isa{q}} \subseteq \{\isa{x}\} \\
& \hspace{10mm} \isand \SPTarget = \sum_{\isa{h} \in \SPSubset} \isa{eval} ~ \Mal{\isa{q}} ~ [\isa{x} \mapsto \isa{h}], 
\\
\rvcheck ~ \SPRandomness_1 ~\isa{rs} & \:\longleftrightarrow\: 
\sumcheck ~ \SPProver ~ \NewSPState ~ (\SPSubset,\: \NewSPPolynomial \: \SPRandomness_1,\: \NewSPTarget \: \SPRandomness_1) ~ \SPRandomness_1 ~ (\isaco{zip} ~ \isa{vs} ~ \isa{rs}),
\end{align*}
where the prover $\SPProver$'s polynomial $\Mal{\isa{q}}$, the prover $\SPProver$'s updated state $\NewSPState$, and the reduced sumcheck instance's polynomial $\NewSPPolynomial$ and target value $\NewSPTarget$ are defined as follows:
\begin{align*}
\hat{\isa{q}} & = \fst ~ (\SPProver ~ (\SPSubset, \SPPolynomial, \SPTarget) ~ \isa{x} ~ \isa{vs}~ \SPRandomness ~ \SPState ), \\
\NewSPState   & = \snd ~ (\SPProver ~ (\SPSubset, \SPPolynomial, \SPTarget) ~ \isa{x} ~ \isa{vs} ~ \SPRandomness ~ \SPState ), \\
\NewSPPolynomial ~ \SPRandomness_1 & = \inst ~ \SPPolynomial ~ [\isa{x}\mapsto \SPRandomness_1], \\
\NewSPTarget  ~ \SPRandomness_1    & = \eval ~ \Mal{\isa{q}} ~ [\isa{x}\mapsto \SPRandomness_1]. 
\end{align*}
Here, $\fst$ and $\snd$ denote a pair's first and second projection respectively. 

We start the proof with the following series of (in)equations, which we explain below.
\begin{align}
    \nonumber
    \begin{split}
    & \Prob_{\SPRandomness_1\cons \isa{rs}}[\:  \\
    & \qquad 
      \sumcheck ~ \SPProver ~ \SPState ~ \SPInstanceTupleIsa ~ \SPRandomness ~ (\isaco{zip} ~ (\isa{x} \cons \isa{vs}) ~ (\SPRandomness_1 \cons \isa{rs})) \\
    & \qquad 
      \isand \failure \:] 
    \end{split} \\
    \label{eq:snd1}
    & = \: \Prob_{\SPRandomness_1\cons \isa{rs}}[\: \verchecks \isand \rvcheck ~ \SPRandomness_1 ~ \isa{rs} \isand  \failure \:] \\
    \label{eq:snd2}
    \begin{split}
    & = \: \Prob_{\SPRandomness_1\cons \isa{rs}}[\: \verchecks \isand \rvcheck ~ \SPRandomness_1 ~ \isa{rs} 
      \isand \failure \isand \Mal{\isa{q}} \neq \isa{q} \:] \\
    \end{split} \\
    \label{eq:snd3}
    \begin{split}
    & \leq \: \Prob_{\SPRandomness_1\cons \isa{rs}}[\: \verchecks \isand \rvcheck ~ \SPRandomness_1 ~ \isa{rs} 
      \isand \Mal{\isa{q}} \neq \isa{q} \:] \\
    \end{split} 
\end{align}

In \eqref{eq:snd1}, we simplify the $\isaco{zip}$ expression to $(\isa{x}, \SPRandomness_1) \cons \isaco{zip} ~ \isa{vs}~ \isa{rs}'$ and unfold the $\sumcheck$ definition. 
To obtain \eqref{eq:snd2}, we split the resulting probability into a sum of two cases, depending on whether or not $\Mal{\isa{q}}  = \isa{q}$ holds, where 
$$
\isa{q} = \sum_{\sigma \,\in\, \substs \: (\isaco{set} \: \isa{vs}) \: \SPSubset} \inst ~ \SPPolynomial ~ \mysigma
$$ 
is the honest prover's polynomial. 
When $\Mal{\isa{q}} = \isa{q}$, the probability is $0$ and hence disappears due to a contradiction between $\failure$ and the evaluation check in $\verchecks$, which we derive using the lemmas \isa{eval\_sum\_inst} and \isa{sum\_merge} from \Cref{subsec:Assumptions}. In~\eqref{eq:snd3}, we over-approximate the remaining case by removing $F$.

Next, we split the probability once more into a sum of two cases, this time depending on whether or not $\eval ~ \hat{\isa{q}} ~ [\isa{x} \mapsto \SPRandomness_1] = \eval ~ \isa{q} ~ [\isa{x} \mapsto \SPRandomness_1]$ holds. 
\begin{align}
    \nonumber
    & \Prob_{\SPRandomness_1\cons \isa{rs}}[\: \verchecks \isand \rvcheck ~ \SPRandomness_1 ~ \isa{rs} \isand \Mal{\isa{q}} \neq \isa{q} \:] \\
    \nonumber
    & = \: \Prob_{\SPRandomness_1\cons \isa{rs}}[\: \verchecks \isand \rvcheck ~ \SPRandomness_1 ~ \isa{rs} \isand \Mal{\isa{q}}  \neq \isa{q}  \isand \\
    \nonumber
    & \hspace{26mm} \eval ~ \Mal{\isa{q}} ~ [\isa{x} \mapsto \SPRandomness_1] = \eval ~ \isa{q} ~ [\isa{x} \mapsto \SPRandomness_1] \:] \\
    \nonumber
    & \, + \: \Prob_{\SPRandomness_1\cons \isa{rs}}[\: \verchecks \isand \rvcheck ~ \SPRandomness_1 ~ \isa{rs} \isand \hat{\isa{q}} \neq \isa{q}  \isand \\
    \nonumber
    & \hspace{26mm} \eval ~ \Mal{\isa{q}} ~ [\isa{x} \mapsto \SPRandomness_1] \neq \eval ~ \isa{q} ~ [\isa{x} \mapsto \SPRandomness_1] \:]\\
    \label{eq:prob-lr}
    & = \: \Probleft + \Probright
\end{align}

The final equation~\eqref{eq:prob-lr} above just introduces the abbreviations $\Probleft$ and $\Probright$ for the two probabilities on its left-hand side. We now bound each of these probabilities separately. By the degree and variable checks in $\verchecks$, the lemma's first assumption, and the $\isa{roots}$ locale assumption, the distinct polynomials $\isa{q}$ and $\Mal{\isa{q}}$ coincide on at most $\SPDegree$ values. Hence, we have 
\begin{align}
\Probleft
          & \leq \frac{\SPDegree}{\card{\UnivA}}.
\end{align}

We derive a bound for the probability $\Probright$ as follows using the induction hypothesis.
\begin{align}
\nonumber 
& \Probright 
\\
\nonumber
& \:\leq\: \Prob_{\SPRandomness_1\cons \isa{rs}}[\: \rvcheck ~ \SPRandomness_1 ~ \isa{rs} \,\isand\, {} \\
\label{ineq:1}
& \hspace{19.5mm}\eval ~ \Mal{\isa{q}} ~ [\isa{x} \mapsto \SPRandomness_1] \neq \eval ~ \isa{q}~ [\isa{x} \mapsto \SPRandomness_1] \:] 
\\
\nonumber
& \:=\: \Prob_{\SPRandomness_1\cons \isa{rs}}[\: \rvcheck ~ \SPRandomness_1 ~ \isa{rs} \,\isand\, {} \\
\label{eq:2}
& \hspace{19.5mm} \NewSPTarget ~ \SPRandomness_1 \:\neq  \sum_{\sigma \,\in\, \isaco{subst} \: (\isaco{set} \: \isa{vs}') \: \SPSubset} \eval ~ (\NewSPPolynomial ~ \SPRandomness_1) ~ \mysigma \:] 
\\
\nonumber
& \:\leq\: \frac{1}{\card{\UnivA}} \sum_{\alpha \in \UnivA} \Prob_{\isa{rs}}[\: \rvcheck ~ \myalpha ~ \isa{rs} \,\isand\, {} \\
\label{ineq:3}
& \hspace{26mm} \NewSPTarget ~ \myalpha \:\neq {\displaystyle \sum_{\sigma \,\in\, \isaco{subst} \: (\isaco{set} \: \isa{vs}') \: \SPSubset} \eval ~ (\NewSPPolynomial ~ \myalpha) ~ \mysigma} \:] 
\\
\label{ineq:4}
& \:\leq\: \frac{1}{\card{\UnivA}} \sum_{\alpha \in \UnivA} \frac{\SPDegree \cdot \isaco{length} ~ \isa{vs}'}{\card{\UnivA}}
\\
\label{ineq:5}
& \:\leq\: \frac{\SPDegree \cdot \isaco{length} ~ \isa{vs}'}{\card{\UnivA}}. 
\end{align}

The inequality~\eqref{ineq:1} holds by monotonicity of $\Prob$. 
The equality~\eqref{eq:2} holds by definition of $(\NewSPTarget ~ \SPRandomness_1)$ and since 
\begin{align*}
\eval ~ \isa{q} ~ [\isa{x} \mapsto \SPRandomness_1] = \sum_{\sigma \,\in\, \isaco{subst} \: (\isaco{set} \: \isa{vs}') \: \SPSubset} \eval ~ (\NewSPPolynomial ~ \SPRandomness_1) ~ \mysigma ,
\end{align*}
by the lemma $\isa{eval\_sum\_inst\_commute}$ from \Cref{subsec:Assumptions}.
To derive \eqref{ineq:3}, we use a lemma about probabilities that allows us to remove $\SPRandomness_1$ from the probability space and to sum over uniformly chosen instantiations with concrete values instead. This step is required for the application of the induction hypothesis.

In~\eqref{ineq:4}, we apply the induction hypothesis which reads as follows: \emph{for all $\SPState$, $\SPPolynomial$, $\SPTarget$, and $\SPRandomness$}, we have 
\begin{align*}
    & \Prob_{\isa{rs} \:\in\: \tuples \, (\isaco{length} \: \isa{vs}'))}[ \\
    & \quad\; \sumcheck ~ \SPProver ~ \SPState ~ (\SPSubset,\; \SPPolynomial,\; \SPTarget) ~ \SPRandomness ~ (\isaco{zip} ~ \isa{vs}' ~ \isa{rs}) \:\land\: {} \\
    & \quad\; 
    \SPTarget \:\neq \sum_{\sigma \: \in \: \substs \: (\isaco{set} \: \isa{vs}') \: \SPSubset} \eval ~ \SPPolynomial ~ \mysigma  \\
    & ] \leq \frac{\SPDegree \cdot \isaco{length} ~ \isa{vs}'}{\card{\UnivA}}, 
\end{align*}
provided that $\vars ~ \SPPolynomial \subseteq \isaco{set} ~ \isa{vs}'$, $\deg ~ \SPPolynomial \leq \SPDegree$, $\distinct ~ \isa{vs}'$ and $\SPSubset \neq \emptyset$. 
We respectively instantiate $\SPState$, $\SPPolynomial$, $\SPTarget$, and $\SPRandomness$ in the induction hypothesis with $\NewSPState$, $(\NewSPPolynomial ~ \myalpha)$, $(\NewSPTarget ~ \myalpha)$, and $\myalpha$ as introduced above. The resulting instantiated assumptions follow from those of the induction step. 
The final simplification step in~\eqref{ineq:5} yields the desired bound for $\Probright$.

We can now continue from~\eqref{eq:prob-lr} to finish the proof of the inductive step as follows.
\begin{align*}
    & \Prob_{\SPRandomness_1\cons \isa{rs}}[\: \verchecks \isand \rvcheck ~ \SPRandomness_1 ~ \isa{rs} \isand \Mal{\isa{q}} \neq \isa{q} \:] 
    \\
    & \: = \: \frac{\SPDegree}{\card{\UnivA}} + \frac{\SPDegree \cdot (\isaco{length} ~ \isa{vs}')}{\card{\UnivA}} 
    \\
    & \: \leq \: \frac{\SPDegree \cdot \isaco{Suc} ~ (\isaco{length} ~ \isa{vs}')}{\card{\UnivA}}.
\end{align*}
This completes the soundness proof.
\end{proof}

\subsection{Differences with Pen-and-Paper Proofs}

Common pen-and-paper proofs of completeness and soundness generally gloss over several details that we had to make precise in our formal proof.

These proofs usually proceed by induction over the arity of the polynomial (cf.~\Cref{app:properties}), ignoring the aforementioned fact that the arity of the polynomial may decrease by more than one in each round. Instead, our proof proceeds by induction on an explicit list of variables which contains all of the variables of the sumcheck polynomial, even if a particular variable never actually appears. Hence, the informal notation $\SPPolynomial(\XVar_1,\ldots,\XVar_\SPArity)$ gives an upper bound on the set of variables that actually occur in $\SPPolynomial$.
In an early stage of our development, we attempted an induction over the arity of $\SPPolynomial$. However, this required stronger assumptions about polynomials that fail to hold for polynomials represented in a unique normal form, as is the case for our instantiation in \Cref{sec:polynomial-instantiation}. For example, we had initially formulated the locale assumptions $\isa{vars\_add}$ and $\isa{vars\_inst}$ from \Cref{fig:locale} with equalities, which we subsequently had to weaken to set inclusions.

Differently from some informal presentations \cite{thaler}, our inductive security analysis uses (constant) polynomials of arity zero as the base case, rather than (univariate) polynomials of arity one. This makes the base case trivial and avoids the repetition of similar reasoning in the base case and inductive steps of the analysis.

Moreover, we had to fully specify all probability spaces involved, another important detail that is often omitted in less formal proofs. This required proving some additional lemmas to reduce the dimension of the probability space in the inductive step and enable the use of the induction hypothesis.

\section{Instantiation with Multivariate Polynomials}
\label{sec:polynomial-instantiation}

Our formalization of the sumcheck protocol in the previous section relies on the abstract functions $\vars$, $\deg$, $\eval$ and $\inst$ and a set of assumptions about them. So far, we have not shown that the assumptions are realizable by any particular mathematical structure. Instantiating the abstract type $\isa{'p}$ with a concrete type of multivariate polynomials (\Cref{sub:multivariate_polynomials}), defining concrete instances of the four functions (\Cref{sub:functions-on-polynomials}), and proving the associated assumptions (\Cref{sub:proving-assumptions}) completes the formalization of the completeness and soundness proofs of the sumcheck protocol for the concrete case of polynomials. Furthermore, this step proves the consistency of our axiomatization and provides confidence in the applicability of our abstract formalization.
To achieve this, we use existing Isabelle/HOL libraries for univariate polynomials~\cite{HOLComputationalAlgebra} and for multivariate polynomials~\cite{Polynomials-AFP}. The latter also connects univariate with multivariate polynomials with at most one variable.

\subsection{The Type of Multivariate Polynomials}
\label{sub:multivariate_polynomials}

The multivariate polynomials of \cite{Polynomials-AFP} rely on the type of polynomial mappings from the Isabelle/HOL library,
which is defined to be isomorphic to the set of functions that are zero ``almost everywhere'', i.e., on all but a finite number of arguments. In Isabelle, this type is defined as follows:
\[
  \isatypedef \isa{'a} \fun_0\! \isa{'b} = \{\isa{f} :: \isa{'a} \fun \isa{'b::zero} \mid \isaco{finite}~\{\isa{x} \mid \isa{f}~\isa{x} \neq 0\}\}
\]
This construction defines a new type from the indicated set of functions and comes with the morphism $\isaco{lookup} :: (\isa{'a} \fun_0\! \isa{'b}) \fun (\isa{'a} \fun \isa{'b})$, which opens the polynomial mapping abstraction and its inverse $\isaco{Abs\_poly\_mapping}$, which creates such an abstraction. The function $\isa{keys} :: (\isa{'a} \fun_0 \isa{'b}) \fun \isa{'a set}$ returns the set of arguments on which the polynomial mapping results in a non-zero value. Polynomial mappings have a rich algebraic structure, and are defined in the theory~\cite{IsabellePolyMapping}. 

We now recap how the authors of~\cite{Polynomials-AFP} construct the types of monomials and then multivariate polynomials using polynomial mappings.
We will use the following running example throughout this subsection to clarify the concepts introduced. The multivariate polynomial 
$$
p(x_1, x_2, x_3) = 3x_1^2x_2x_3 + 2x_1x_3 + x_3^2
$$ 
has three variables and three \emph{monomials} with non-zero coefficients, namely, $m_1 = x_1^2x_2x_3$ with coeffcient $3$, $m_2 = x_1x_3$ with coefficient $2$, and $m_3 = x_3^2$ with coefficient $1$.

First, we can use polynomial mappings of type $\isa{nat} \fun_0 \isa{nat}$ to represent monomials by mapping variables to their exponent. For example, the monomial $m_1$ would be represented as $\isaco{Abs\_poly\_mapping}~((\lambda \isa{x}.\,0)(1 \mapsto 2, 2 \mapsto 1, 3 \mapsto 1))$, which we abbreviate to $[1 \mapsto 2, 2 \mapsto 1, 3 \mapsto 1]$. 
The type of multivariate polynomials then maps monomials of type $\isa{nat} \fun_0 \isa{nat}$ to their coefficients of type \isa{'a} and is defined by 
$$
\isatypedef \isa{'a } \isa{mpoly} = \isa{UNIV} :: ((\isa{nat} \fun_0 \isa{nat}) \fun_0 \isa{'a})~\isa{set}.
$$
Here, $\isa{UNIV}$ is the set of all elements of the indicated type. This type definition comes with the morphisms $\isaco{mapping\_of}$, which opens the type $\isa{'a mpoly}$, returning the underlying polynomial mapping of type $(\isa{nat} \fun_0 \isa{nat}) \fun_0 \isa{'a}$.
Accordingly, the polynomial $p(x_1, x_2, x_3)$ can be represented by $[m_1 \mapsto 3, m_2 \mapsto 2, m_3 \mapsto 1]$, i.e., mapping all other monomials to 0. Note that $p$ in this example is of type \isa{nat mpoly}.

\subsection{Functions on Multivariate Polynomials}
\label{sub:functions-on-polynomials}

We now define the functions $\vars$, $\deg$, $\eval$ and $\inst$ to instantiate our locale. While we defined polynomial instantiation ourselves, the former three are already defined in~\cite{Polynomials-AFP}.

\subsubsection{Variables} 
The variable function for a multivariate polynomial is defined by
\begin{align*}
    \isadefinition &\varsco :: \isa{'a mpoly} \fun \isa{nat set} \isawhere \\
    & \varsco ~ \isa{p} = \bigcup_{m \,\in\, \isaco{keys} \: (\isaco{mapping\_of} \: \isa{p} )} (\isaco{keys} ~ \isa{m}).
\end{align*}
Here, $\isaco{keys} ~ (\isaco{mapping\_of} ~ \isa{p})$ is the set of the polynomial $\isa{p}$'s monomials and the union of these monomials' sets of keys constitutes $\isa{p}$'s variables. We use this function to instantiate our variables function in the locale. With our running example, we have $\varsco ~ \isa{p} = \{1, 2, 3\}$, representing the variables $x_1, x_2, x_3$.

\subsubsection{Degree} 
As our degree function, $\deg$, we use 
\begin{align*}
    & \isaliftdefinition \isaco{total\_degree} :: \isa{'a mpoly} \fun \isa{nat} ~ \isais \\
    & \quad \lambda p. \: \isaco{Max} ~ (\{0\} \cup \{ \sum_{\isa{v} \,\in\, \isaco{keys} \: \isa{m}} \isaco{lookup} ~ \isa{m} ~ \isa{v} \mid \isa{m} \in \isaco{keys} ~ \isa{p}\}).
\end{align*}
This function returns the maximum of any monomial's sum of exponents in the polynomial. More precisely, the expression $(\isaco{keys} ~ \isa{p})$ denotes the set of $\isa{p}$'s monomials and the sum adds up the exponents of each of the monomial $\isa{m}$'s variables $\isa{x}$. 
This definition makes use of Isabelle's lifting definition mechanism, which allows one to write definitions in terms of the concrete representations of a newly defined type (such as $\isa{'a mpoly}$) and automatically inserts the required morphisms. Here, the $\lambda$-abstracted $\isa{p}$ is of type $((\isa{nat} \fun_0 \isa{nat}) \fun \isa{'a}) \fun \isa{nat}$, which is automatically lifted to the argument type $\isa{'a mpoly}$ of the defined function using the morphism $\isaco{mapping\_of}$.
For our running example, we thus get $\isaco{total\_degree} ~ \isa{p} = 4$ . We note that we apply this function only to multivariate polynomials in at most one variable, where the definition collapses to the expected one. 

\subsubsection{Evaluation} 
We use the existing function $\isaco{insertion}$, which is defined as follows.
\begin{align*}
    \isadefinition &\isaco{insertion\_fun} :: (\isa{nat}\fun\isa{'a}) \fun\\
    & \hspace{34mm}((\isa{nat} \Rightarrow_0 \isa{nat}) \fun \isa{'a}) \fun \isa{'a} \\
    \isawhere & \isaco{insertion\_fun } ~ \mysigma ~ \isa{p} = \\
    &(\sum_{\isa{m} \,\in\, \isaco{keys} \: \isa{p}}  \isa{p} ~ \isa{m} \cdot (\prod_{\isa{v} \,\in\, \isaco{keys} \: \isa{m}} (\mysigma ~ \isa{v})^{\isaco{lookup} \: \isa{m} \: \isa{v}}))
\end{align*}
This definition, through lifting, yields the $\isaco{insertion}$ function of type $(\isa{nat} \fun \isa{'a}) \fun \isa{'a mpoly} \fun \isa{'a}$. We then define our equivalent evaluation function $\evalco :: \isa{'a mpoly} \fun (\isa{nat}, \isa{'a}) ~ \isa{subst} \fun \isa{'a}$ by swapping its arguments and converting our partial substitution into a total one (with some irrelevant undefined values).
Taking the substitution $\mysigma = [1 \mapsto 3, 2 \mapsto 1, 3 \mapsto 2]$ in our running example $p(x_1, x_2, x_3) = 3x_1^2x_2x_3 + 2x_1x_3 + x_3^2$ replaces $x_1$ by $3$, $x_2$ by $1$ and $x_3$ by $2$ in $p$. Therefore,  $\eval ~ p ~ \mysigma = 3 \cdot 3^2\cdot 1\cdot2 + 2\cdot 3 \cdot 2 + 2^2 = 70$.

\subsubsection{Instantiation} 
Polynomial instantiation substitutes certain variables with values to obtain a polynomial of smaller arity. It can be understood as a partial evaluation. In contrast to evaluation, not all variables need to be instantiated. However, instantiating all variables is equivalent to evaluation. 
For our example polynomial $p(x_1, x_2, x_3) = 3x_1^2x_2x_3 + 2x_1x_3 + x_3^2$ and the substitution $\mysigma = [1 \mapsto 3, 2 \mapsto 1]$, which instantiates $x_1$ and $x_2$ but not $x_3$, the resulting polynomial is $33x_3 + x_3^2$. We will discuss this example in more detail, while introducing the definitions. 

We first define instantiation on monomials and then extend it to polynomials. 
Recall that the monomials of $p$ are $x_1^2x_2x_3$, $x_1x_3$ and $x_3^2$. Applying the substitution to each monomial independently yields $27x_3$, $6x_3$ and $x_3^2$. We note that each of these terms has a coefficient and a residual monomial, which we respectively define in the functions $\isaco{inst\_mon\_coeff}$ and $\isaco{inst\_mon\_resid}$ as follows. 
\begin{align*}
    \isaliftdefinition &\isaco{inst\_mon\_coeff} ::  \\
    &(\isa{nat}\fun_0\isa{nat}) \fun (\isa{nat}, \isa{'a})~\isa{subst} \fun \isa{'a} \\
    \isais & \lambda \isa{m} \: \mysigma. \: (\prod_{\isa{v} \,\in\, \isa{dom}\:\sigma} (\mysigma ~ \isa{v})^{\isa{m} \: \isa{v}}) \\
    \isaliftdefinition &\isaco{inst\_mon\_resid :: }\\
    & (\isa{nat} \fun_0 \isa{nat}) \fun (\isa{nat}, \isa{'a})~\isa{subst} \fun (\isa{nat}\fun_0\isa{nat}) \\
    \isais & \lambda \isa{m} \: \mysigma \: \isa{v}. \: \isaif \isa{v} \notin \isa{dom} ~ \mysigma \isathen \isa{m} ~ \isa{v} \isaelse 0 
\end{align*}

Once we have applied the substitution to each monomial, it remains to multiply these by their original coefficients in $\isa{p}$ and group them by the residual monomials. This is what the function $\isaco{inst\_fun}$ will do, which in our example simplifies $3 \cdot 9x_3 + 2 \cdot 3x_3 + x_3^2$ to the desired $33x_3 + x_3^2$. The function is defined as follows:
\begin{align*}
    &\isadefinition \isaco{inst\_fun} :: ((\isa{nat}\fun_0\isa{nat}) \fun \isa{'a}) \fun  \\
    & \phantom{\isadefinition \isaco{inst\_fun} :: {}} 
      (\isa{nat}, \isa{'a})~\isa{subst} \fun (\isa{nat} \fun_0 \isa{nat}) \fun \isa{'a} \\
    & \!\!\!\isawhere\\
    & \quad \isaco{inst\_fun} ~ \isa{p} ~ \mysigma =  \\
    & \qquad (\lambda \isa{m}. \: \sum_{\isa{m}' \,\mid\, \isaco{inst\_mon\_resid} \: \isa{m}' \: \sigma \,=\, \isa{m}} \isa{p} ~ \isa{m}' \cdot \isaco{inst\_mon\_coeff} ~ \isa{m}' \: \mysigma). 
\end{align*}
Finally, we obtain the function $\instco$ of the correct type $\isa{'a mpoly} \fun (\isa{nat}, \isa{'a}) ~ \isa{subst} \fun \isa{'a mpoly}$ through lifting.

\subsection{Proving the Assumptions}
\label{sub:proving-assumptions}

We now prove that the locale assumptions from \cref{subsec:Assumptions} hold for the functions $\varsco$, $\degco$, $\evalco$, and $\instco$ on polynomials. Note that we discuss the assumptions relating to $\inst$ separately, although in \Cref{fig:locale} they are grouped with those for $\vars$, $\deg$ and $\eval$.

 \subsubsection{Variables Assumptions} 
 There are existing library lemmas for the function $\varsco$ that exactly correspond to our assumptions $\isa{vars\_finite}$ and $\isa{vars\_add}$, and the proof of $\isa{vars\_zero}$ follows directly from the definitions of the variable function and the zero polynomial. 

\subsubsection{Degree Assumptions} 
We must prove the assumptions $\isa{deg\_zero}$ and $\isa{deg\_add}$ for the total degree function $\degco$. The first one is covered by a library lemma. We establish $\isa{deg\_add}$ separately. Its proof uses the definition of total degree and requires simple properties about the $\isaco{Max}$ function and the addition of multivariate polynomials. 

\subsubsection{Evaluation Assumptions} 
We prove the assumptions $\isa{eval\_zero}$ and $\isa{eval\_add}$ directly using existing library lemmas about the $\isaco{insertion}$ function and the definition of $\evalco$. 

\subsubsection{Instantiation Assumptions}
The proofs of the instantiation assumptions $\isa{vars\_inst}$, $\isa{deg\_inst}$ and $\isa{eval\_inst}$ required more work since the concept of polynomial instantiation did not exist in the existing theories.  Although the proofs were more elaborate than for the other assumptions, they did not require more than basic theorems about sums, maps, and finiteness, along with the definitions of monomial and polynomial instantiation.

\subsubsection{Roots Assumption} 
\label{par:roots_assumption}
We prove the roots assumption using a lemma stating that the number of roots of a univariate polynomial (of type $\isa{'a mpoly}$) is bounded by its total degree (which coincides with the expected degree for such polynomials). We first establish this result for a specialized type, $\isa{'a poly}$, of univariate polynomials, which exists in the Isabelle/HOL library, and then transfer it to univariate polynomials of type $\isa{'a mpoly}$. We explain the type $\isa{'a poly}$ in \Cref{app:univariate-polys} and present the proof of the roots bound in \Cref{app:roots_bound}. 

\section{Conclusion and Future Work}
\label{sec:conclusions}

In this paper, we formally verified the completeness and soundness properties of the sumcheck protocol. Our work took place in three stages:
\begin{itemize}
    \item Producing a new pen-and-paper security analysis of the sumcheck protocol with enough detail for a machine-checkable analysis;
    \item Distilling the necessary properties of multivariate polynomials used in the sumcheck protocol, and proving the security of a generalized sumcheck protocol based on these properties in Isabelle; and
    \item Showing that existing Isabelle formalizations of multivariate polynomials satisfy the necessary properties, leading to a full security analysis of the classic sumcheck protocol.
\end{itemize}

The sumcheck protocol is widely used, with applications ranging from computational complexity theory to cryptography. We hope that the contrast between the apparent straightforwardness of the well-known pen-and-paper proofs for the sumcheck protocol and the thoroughness of our formalization motivates the use of formal verification. Formalizing proofs will be crucial to prevent errors and ensure security as cryptographic systems become increasingly complex.

We also hope that our formalization helps to produce machine-checkable security analyses of other recursive protocols in the future, whether they are based on the sumcheck protocol, or other probabilistic proof systems.
Future extensions of our work would be to create machine-checkable proofs of statements in complexity theory relying on the sumcheck protocol \cite{introSumcheck,IPisPSPACE,shen1992ip,Jawale2020SNARGsFB,ChoudhuriHKPRR19,KalaiLV23}, analyses of protocols such as the GKR protocol~\cite{ref1auro}, which rely directly on sumcheck, generalisations of the sumcheck protocol to cryptographic groups \cite{scarguments} and tensor codes \cite{tensorsumcheck}, and producing verified implementations of these protocols using the Isabelle code generator. Setting up the code generator to extract an implementation of our verified sumcheck protocol should be feasible with limited additional effort given the extensive support for executable polynomials in~\cite{Polynomials-AFP}.


\section*{Acknowledgments}
\label{sec:acknowledgements}

We would like to thank the anonymous reviewers, the shepherd, and David Basin for their useful feedback, which helped improving the paper.
The first author received the support of a fellowship (code LCF/BQ/EU20/11810060) from La Caixa Foundation (ID 100010434).

\printbibliography

@article{IshaiKOS09,
  author       = {Ishai, Yuval and Kushilevitz, Eyal and Ostrovsky, Rafail and Sahai, Amit},
  title        = {Zero-Knowledge Proofs from Secure Multiparty Computation},
  journal      = {{SIAM} J. Comput.},
  volume       = {39},
  pages        = {1121--1152},
  year         = {2009},
}

@Misc{gnark,
  title = {Consensys gnark Library},
  url   = {https://github.com/ConsenSys/gnark},
}

@article{SchwartzZippel-AFP,
  author  = {Sunpill Kim and Yong Kiam Tan},
  title   = {The {S}chwartz-{Z}ippel Lemma},
  note    = {\url{https://isa-afp.org/entries/Schwartz_Zippel.html}, Formal proof development},
  issn    = {2150-914x},
  journal = {Archive of Formal Proofs},
  month   = {04},
  year    = {2023},
}

@Misc{Iden3,
  title = {Iden3 snarkjs Library},
  url   = {https://github.com/iden3/snarkjs},
}

@Misc{Dusk,
  title = {Dusk Network Plonk Implementation},
  url   = {https://github.com/dusk-network/plonk},
}

@Misc{SecBit,
  title = {{SecBit} Labs Zero-Knowledge Proofs Toolkit},
  url   = {https://github.com/sec-bit/ckb-zkp},
}

@Misc{ZenGo,
  title = {{ZenGo} Zero Knowledge {P}aillier Implementation},
  url   = {https://github.com/ZenGo-X/zk-paillier},
}

@Misc{HOLComputationalAlgebra,
  title = {{Isabelle/HOL Computational Algebra Library}},
  url   = {https://isabelle.in.tum.de/library/HOL/HOL-Computational_Algebra}
}

@Misc{SWB19,
  author = {Swihart, Josh and Winston, Benjamin and Bowe, Sean},
  title  = {ZCash Counterfeiting Vulnerability Successfully Remediated},
  url    = {https://electriccoin.co/blog/zcash-counterfeiting-vulnerability-successfully-remediated/},
  year   = {2019},
}

@Misc{Miller22,
  author = {Miller, Jim},
  title  = {Coordinated disclosure of vulnerabilities affecting {Girault}, {Bulletproofs}, and {PlonK}},
  url    = {https://blog.trailofbits.com/2022/04/13/part-1-coordinated-disclosure-of-vulnerabilities-affecting-girault-bulletproofs-and-plonk/},
  year   = {2022},
}

@inproceedings{Ben-SassonCTV14,
  author       = {Ben--Sasson, Eli and Chiesa, Alessandro and Tromer, Eran and Virza, Madars},
  title        = {Succinct Non-Interactive Zero Knowledge for a von Neumann Architecture},
  booktitle    = {Proceedings of the 23rd {USENIX} Security Symposium},
  pages        = {781--796},
  series    = {{USENIX}~'14},
  year         = {2014},
}

@Misc{BunzF22,
  author       = {B{\"{u}}nz, Benedikt and Fisch, Ben},
  title        = {Schwartz-{Z}ippel for multilinear polynomials mod {N}},
  howpublished = {{IACR} Cryptology ePrint Archive, Report 2022/458},
  url          = {https://eprint.iacr.org/2022/458},
  year         = {2022},
}

@inproceedings{DamgardF02,
  author       = {Damg{\aa}rd, Ivan and Fujisaki, Eiichiro},
  title        = {A Statistically-Hiding Integer Commitment Scheme Based on Groups with
                  Hidden Order},
  booktitle    = {Proceedings of the 8th International Conference on the Theory and Application of Cryptology and Information Security},
  series       = {{ASIACRYPT}~'02},
  pages        = {125--142},
  year         = {2002},
}

@inproceedings{FujisakiO97,
  author       = {Fujisaki, Eiichiro and Okamoto, Tatsuaki},
  title        = {Statistical Zero Knowledge Protocols to Prove Modular Polynomial Relations},
  booktitle    = {Proceedings of the 17th Annual International Cryptology Conference},
  series       = {{CRYPTO}~'97},
  volume       = {1294},
  pages        = {16--30},
  year         = {1997},
}

@Article{tensorsumcheck,
  author  = {Meir, Or},
  title   = {{IP} = {PSPACE} using error-correcting codes},
  pages   = {380--403},
  volume  = {42},
  journal = {SIAM Journal on Computing},
  year    = {2013},
}

@InProceedings{introductionIP,
  author    = {Goldwasser, Shafi and Micali, Silvio and Rackoff, Charles},
  booktitle = {Proceedings of the 17th Annual {ACM} Symposium on Theory of Computing},
  title     = {The Knowledge Complexity of Interactive Proof-Systems (Extended Abstract)},
  pages     = {291--304},
  series    = {STOC~'85},
  year      = {1985},
}

@InProceedings{eqintroductionIP,
  author    = {Babai, L{\'{a}}szl{\'{o}}},
  booktitle = {Proceedings of the 17th Annual {ACM} Symposium on Theory of Computing},
  title     = {Trading Group Theory for Randomness},
  pages     = {421--429},
  series    = {{STOC}~'85},
  year      = {1985},
}

@Article{IPisPSPACE,
  author  = {Shamir, Adi},
  title   = {\uppercase{IP = PSPACE}},
  number  = {4},
  pages   = {869–877},
  volume  = {39},
  journal = {J. ACM},
  month   = {10},
  year    = {1992},
}

@Article{introSumcheck,
  author  = {Lund, Carsten and Fortnow, Lance and Karloff, Howard and Nisan, Noam},
  title   = {Algebraic Methods for Interactive Proof Systems},
  number  = {4},
  pages   = {859–868},
  volume  = {39},
  journal = {J. ACM},
  year    = {1992},
}

@misc{CanettiCHLRR18,
  author    = {Canetti, Ran and Chen, Yilei and Holmgren, Justin and Lombardi, Alex and Rothblum, Guy N. and Rothblum, Ron D.},
  title     = {{F}iat--{S}hamir From Simpler Assumptions},
  howpublished = {Cryptology ePrint Archive, Report 2018/1004},
  year      = {2018},
}

@InProceedings{Jawale2020SNARGsFB,
  author    = {Jawale, Ruta and Kalai, Yael Tauman and Khurana, Dakshita and Zhang, Rachel Yun},
  booktitle = {Proceedings of the 53rd Annual {ACM} {SIGACT} Symposium on Theory of Computing},
  title     = {{SNARG}s for bounded depth computations and {PPAD} hardness from sub-exponential {LWE}},
  pages     = {708--721},
  series    = {{STOC}~'21},
  year      = {2021},
}

@Misc{thaler,
  author = {Justin Thaler},
  title  = {The Sum-Check Protocol},
  note   = {\url{https://people.cs.georgetown.edu/jthaler/sumcheck.pdf}},
}

@Article{ref1auro,
  author  = {Goldwasser, Shafi and Kalai, Yael Tauman and Rothblum, Guy N},
  title   = {Delegating computation: interactive proofs for muggles},
  number  = {4},
  pages   = {1--64},
  volume  = {62},
  journal = {Journal of the ACM},
  year    = {2015},
}

@InProceedings{tensor1,
  author    = {Bootle, Jonathan and Chiesa, Alessandro and Groth, Jens},
  booktitle = {Proceedings of the 18th Theory of Cryptography Conference},
  title     = {Linear-Time Arguments with Sublinear Verification from Tensor Codes},
  pages     = {19--46},
  series    = {{TCC}~'20},
  year      = {2020},
}

@article{bellare2004code,
  title={Code-based game-playing proofs and the security of triple encryption},
  author={Bellare, Mihir and Rogaway, Phillip},
  journal={Cryptology ePrint Archive},
  year={2004}
}

@article{cryptholsigmacommitments,
  title={Formalising $\Sigma$-Protocols and Commitment Schemes using {CryptHOL}},
  author={Butler, David and Lochbihler, Andreas and Aspinall, David and Gasc{\'o}n, Adri{\`a}},
  journal={Journal of Automated Reasoning},
  volume={65},
  number={4},
  pages={521--567},
  year={2021},
  publisher={Springer}
}

@InProceedings{MPCformalization,
  author    = {Sidorenco, Nikolaj and Oechsner, Sabine and Spitters, Bas},
  booktitle = {2021 IEEE 34th Computer Security Foundations Symposium (CSF)},
  title     = {Formal security analysis of {MPC}-in-the-head zero-knowledge protocols},
  pages     = {1--14},
  series    = {CSF~'21},
  year      = {2021},
}

@article{Polynomials-AFP,
  author  = {Christian Sternagel and René Thiemann and Alexander Maletzky and Fabian Immler and Florian Haftmann and Andreas Lochbihler and Alexander Bentkamp},
  title   = {Executable Multivariate Polynomials},
  note    = {\url{https://isa-afp.org/entries/Polynomials.html}, Formal proof development},
  issn    = {2150-914x},
  journal = {Archive of Formal Proofs},
  month   = {8},
  year    = {2010},
}

@InProceedings{dilithium_formalization,
  author    = {Barbosa, Manuel and Barthe, Gilles and Doczkal, Christian and Don, Jelle and Fehr, Serge and Gr{\'{e}}goire, Benjamin and Huang, Yu{-}Hsuan and H{\"{u}}lsing, Andreas and Lee, Yi and Wu, Xiaodi},
  booktitle = {Proceedings of the 43rd Annual International Cryptology Conference},
  title     = {Fixing and Mechanizing the Security Proof of {Fiat-Shamir} with Aborts and {D}ilithium},
  pages     = {358--389},
  series    = {{CRYPTO}~'23},
  year      = {2023},
}

@InProceedings{darkfix2,
  author    = {Arun, Arasu and Ganesh, Chaya and Lokam, Satya V. and Mopuri, Tushar and Sridhar, Sriram},
  booktitle = {Proceedings of the 26th {IACR} International Conference on Practice and Theory of Public-Key Cryptography},
  title     = {Dew: {A} Transparent Constant-Sized Polynomial Commitment Scheme},
  pages     = {542--571},
  series    = {{PKC}~'23},
  year      = {2023},
}

@InProceedings{bootlefix1,
  author    = {Attema, Thomas and Cramer, Ronald and Kohl, Lisa},
  booktitle = {Proceedings of the 41st Annual International Cryptology Conference},
  title     = {A Compressed Sigma-Protocol Theory for Lattices},
  pages     = {549--579},
  series    = {{CRYPTO}~'21},
  year      = {2021},
}

@InProceedings{bootlefix2,
  author    = {Albrecht, Martin R. and Lai, Russell W. F.},
  booktitle = {Proceedings of the 41st Annual International Cryptology Conference},
  title     = {Subtractive Sets over Cyclotomic Rings - Limits of {S}chnorr-Like Arguments over Lattices},
  pages     = {519--548},
  series    = {{CRYPTO}~'21},
  year      = {2021},
}

@InProceedings{CAC,
  author    = {Barbosa, Manuel and Barthe, Gilles and Bhargavan, Karthik and Blanchet, Bruno and Cremers, Cas and Liao, Kevin and Parno, Bryan},
  booktitle = {Proceedings of the 42nd {IEEE} Symposium on Security and Privacy},
  title     = {{SoK}: Computer-Aided Cryptography},
  pages     = {777--795},
  series    = {{S\&P}~'21},
  year      = {2021},
}

@Article{shen1992ip,
  author  = {Shen, Alexander},
  title   = {{IP} = {PSPACE}: simplified proof},
  number  = {4},
  pages   = {878--880},
  volume  = {39},
  journal = {Journal of the ACM},
  year    = {1992},
}

@Misc{BoltonBailey,
  author       = {Bailey, Bolton},
  title        = {Formalization of {SNARKs}},
  howpublished = {\url{https://github.com/BoltonBailey/formal-snarks-project}},
  year         = {last accessed 22.03.2023},
}

@InProceedings{gemini,
  author    = {Bootle, Jonathan and Chiesa, Alessandro and Hu, Yuncong and Orr{\`{u}}, Michele},
  booktitle = {Proceedings of the 41st Annual International Conference on the Theory and Applications of Cryptographic Techniques},
  title     = {Gemini: Elastic SNARKs for Diverse Environments},
  pages     = {427--457},
  series    = {{EUROCRYPT}~'22},
  year      = {2022},
}

@Misc{IsabellePolyMapping,
  author  = {Lochbihler, Andreas and Haftmann, Florian},
  title   = {Theory {HOL}-Library.{P}oly\_{M}apping},
  url     = {https://isabelle.in.tum.de/library/HOL/HOL-Library/Poly_Mapping.html},
  urldate = {2023-08-17},
}

@Misc{IsabellePolynomial,
  author  = {Huffman, Brian and Ballarin, Clemens and Chaieb, Amine and Haftmann, Florian},
  title   = {Theory Polynomial},
  url     = {https://isabelle.in.tum.de/library/HOL/HOL-Computational_Algebra/Polynomial.html},
  urldate = {2023-08-17},
}

@Misc{BM23,
  author       = {Bailey, Bolton and Miller, Andrew},
  title        = {Formalizing Soundness Proofs of {SNARKs}},
  howpublished = {IACR Cryptology ePrint Archive, Report 2023/656},
  year         = {2023},
}

@inproceedings{GennaroGP013,
  author       = {Gennaro, Rosario and Gentry, Craig and Parno, Bryan and Raykova, Mariana},
  title        = {Quadratic Span Programs and Succinct {NIZK}s without {PCP}s},
  booktitle    = {Proceedings of the 32nd Annual International  Conference on the Theory and Applications of Cryptographic Techniques},
  series       = {EUROCRYPT~'13'},
  pages        = {626--645},
  year         = {2013},
}

@InProceedings{ParnoHG013,
  author    = {Parno, Bryan and Howell, Jon and Gentry, Craig and Raykova, Mariana},
  booktitle = {Proceedings of the 2013 {IEEE} Symposium on Security and Privacy},
  title     = {Pinocchio: Nearly Practical Verifiable Computation},
  pages     = {238--252},
  series    = {S\&P~'13},
  year      = {2013},
}

@inproceedings{Groth16,
  author       = {Groth, Jens},
  title        = {On the Size of Pairing-Based Non-Interactive Arguments},
  booktitle    = {Proceedings of the 35th Annual International Conference on the Theory and Applications of Cryptographic Techniques},
  series       = {EUROCRYPT~'16},
  pages        = {305--326},
  year         = {2016},
}

@InProceedings{BagheryKSV21,
  author    = {Baghery, Karim and Kohlweiss, Markulf and Siim, Janno and Volkhov, Mikhail},
  booktitle = {Proceedings of the 25th International Conference on Financial Cryptography and Data Security},
  title     = {Another Look at Extraction and Randomization of {Groth}'s {zk-SNARK}},
  pages     = {457--475},
  series    = {{FC}~'21},
  year      = {2021},
}

@Misc{Lipmaa19,
  author       = {Lipmaa, Helger},
  title        = {Simulation-Extractable {SNARKs} Revisited},
  howpublished = {{IACR} Cryptology ePrint Archive, Report 2019/612},
  url          = {https://eprint.iacr.org/2019/612},
  year         = {2022},
}

@Misc{babysnark,
  title = {BabySNARK},
  url   = {https://github.com/initc3/babySNARK},
}

@InProceedings{FU22,
  author    = {Firsov, Denis and Unruh, Dominique},
  booktitle = {Proceedings of the 36th {IEEE} Computer Security Foundations Symposium},
  title     = {Zero-Knowledge in {EasyCrypt}},
  pages     = {1--16},
  series    = {{CSF}~'23},
  year      = {2023},
}

@article{AFOU23,
  author       = {Almeida, Jos{\'e} Bacelar and Firsov, Denis and Oliveira, Tiago and Unruh, Dominique},
  title        = {{Schnorr} protocol in {Jasmin}},
  howpublished = {{IACR} Cryptology ePrint Archive, Report 2023/752},
  year         = {2023},
  url          = {https://eprint.iacr.org/2023/752.pdf},
}

@InProceedings{ABEGPP21,
  author    = {Almeida, Jos{\'{e}} Bacelar and Barbosa, Manuel and Correia, Manuel L. and Eldefrawy, Karim and Graham{-}Lengrand, St{\'{e}}phane and Pacheco, Hugo and Pereira, Vitor},
  booktitle = {Proceedings of the 2021 {ACM} {SIGSAC} Conference on Computer and Communications Security},
  title     = {Machine-checked {ZKP} for {NP} relations: Formally Verified Security Proofs and Implementations of {MPC}-in-the-Head},
  series    = {{CCS}~'21},
}

@InProceedings{streamingIPs,
  author    = {Cormode, Graham and Mitzenmacher, Michael and Thaler, Justin},
  booktitle = {Proceedings of the 3rd Innovations in Theoretical Computer Science Conference},
  title     = {Practical Verified Computation with Streaming Interactive Proofs},
  pages     = {90--112},
  series    = {ITCS~'12},
  year      = {2012},
}

@InProceedings{spartan,
  author    = {Setty, Srinath},
  booktitle = {Proceedings of the 40th Annual International Cryptology Conference},
  title     = {Spartan: Efficient and General-Purpose {zkSNARKs} Without Trusted Setup},
  pages     = {704--737},
  series    = {CRYPTO~'20},
  year      = {2020},
}

@inproceedings{BootleCCGP16,
  author    = {Bootle, Jonathan and Cerulli, Andrea and Chaidos, Pyrros and Groth, Jens and Petit, Christophe},
  title     = {Efficient Zero-Knowledge Arguments for Arithmetic Circuits in the Discrete Log Setting},
  booktitle = {Proceedings of the 35th Annual International Conference on Theory and Application of Cryptographic Techniques},
  series    = {EUROCRYPT~'16},
  pages     = {327--357},
  year      = {2016},
}

@InProceedings{BlockHRRS20,
  author    = {Block, Alexander R. and Holmgren, Justin and Rosen, Alon and Rothblum, Ron D. and Soni, Pratik},
  booktitle = {Proceedings of the 18th Theory of Cryptography Conference},
  title     = {Public-Coin Zero-Knowledge Arguments with (almost) Minimal Time and Space Overheads},
  pages     = {168--197},
  series    = {{TCC}~'20},
  year      = {2020},
}

@InProceedings{BuenzBBPWM18,
  author    = {B\"unz, Benedikt and Bootle, Jonathan and Boneh, Dan and Poelstra, Andrew and Wuille, Pieter and Maxwell, Greg},
  booktitle = {Proceedings of the 39th IEEE Symposium on Security and Privacy},
  title     = {Bulletproofs: Short Proofs for Confidential Transactions and More},
  pages     = {315--334},
  series    = {{S\&P}~'18},
  year      = {2018},
}

@InProceedings{darksnark,
  author    = {B{\"{u}}nz, Benedikt and Fisch, Ben and Szepieniec, Alan},
  booktitle = {Proceedings of the 39th Annual International Conference on the Theory and Applications of Cryptographic Techniques},
  title     = {Transparent {SNARKs} from {DARK} Compilers},
  pages     = {677--706},
  series    = {{EUROCRYPT}~'20},
  year      = {2020},
}

@InProceedings{bootleZK,
  author    = {Bootle, Jonathan and Lyubashevsky, Vadim and Nguyen, Ngoc Khanh and Seiler, Gregor},
  booktitle = {Proceedings of the 40th Annual International Cryptology Conference},
  title     = {A Non-{PCP} Approach to Succinct Quantum-Safe Zero-Knowledge},
  pages     = {441--469},
  series    = {CRYPTO~'20},
  year      = {2020},
}

@InProceedings{tensor3,
  author    = {Bootle, Jonathan and Chiesa, Alessandro and Liu, Siqi},
  booktitle = {Proceedings of the 42nd Annual International Conference on Theory and Application of Cryptographic Techniques},
  title     = {Zero-Knowledge Succinct Arguments with a Linear-Time Prover},
  pages     = {275--304},
  series    = {EUROCRYPT~'22},
  year      = {2022},
}

@inproceedings{RonZewiR22,
  author    = {{Ron-Zewi}, Noga and Rothblum, Ron D.},
  booktitle = {Proceedings of the 54th Annual ACM Symposium on Theory of Computing},
  title     = {Proving as Fast as Computing: Succinct Arguments with Constant Prover Overhead},
  series    = {STOC~'22},
  year      = {2022},
}

@InProceedings{HolmgrenR22,
  author    = {Holmgren, Justin and Rothblum, Ron},
  booktitle = {Proceedings of the 42nd Annual International Cryptology Conference},
  title     = {Faster Sounder Succinct Arguments and {IOP}s},
  pages     = {474--503},
  series    = {CRYPTO~'22},
  year      = {2022},
}

@InProceedings{darkfix1,
  author    = {Block, Alexander R and Holmgren, Justin and Rosen, Alon and Rothblum, Ron D and Soni, Pratik},
  booktitle = {Proceedings of the 41st Annual International Cryptology Conference},
  title     = {Time- and Space-Efficient Arguments from Groups of Unknown Order},
  pages     = {123--152},
  series    = {{CRYPTO}~'21},
  year      = {2021},
}

@InProceedings{scarguments,
  author    = {Bootle, Jonathan and Chiesa, Alessandro and Sotiraki, Katerina},
  booktitle = {Proceedings of the 41st Annual International Cryptology Conference},
  title     = {Sumcheck Arguments and their Applications},
  note      = {Extended version at \url{https://eprint.iacr.org/2021/333.pdf}.},
  pages     = {681--710},
  series    = {{CRYPTO}~'21},
  year      = {2021},
}

@inproceedings{ChoudhuriHKPRR19,
  author       = {Rai Choudhuri, Arka and
                  Hub{\'{a}}cek, Pavel and
                  Kamath, Chethan and
                  Pietrzak, Krzysztof and
                  Rosen, Alon and
                  Rothblum, Guy N.},
  title        = {Finding a {N}ash equilibrium is no easier than breaking {F}iat-{S}hamir},
  booktitle    = {Proceedings of the 51st Annual {ACM} {SIGACT} Symposium on Theory of Computing},
  pages        = {1103--1114},
  series    = {{STOC}~'19},
  year         = {2019},
}

@inproceedings{KalaiLV23,
  author       = {Tauman Kalai, Yael and
                  Lombardi, Alex and
                  Vaikuntanathan, Vinod},
  title        = {{SNARGs} and {PPAD} Hardness from the Decisional {D}iffie-{H}ellman Assumption},
  booktitle    = {Proceedings of the 42nd Annual International
                  Conference on the Theory and Applications of Cryptographic Techniques},
  series       = {{EUROCRYPT}~'23},
  pages        = {470--498},
  year         = {2023},
}

@Book{NipkowPW02,
  author = {Tobias Nipkow and Lawrence C. Paulson and Markus Wenzel},
  title  = {Isabelle/HOL - {A} Proof Assistant for Higher-Order Logic},
  series = {Lecture Notes in Computer Science},
  url    = {https://doi.org/10.1007/3-540-45949-9},
  volume = {2283},
  year   = {2002},
}

@article{Thaler2022,
	author = {Justin Thaler},
	doi = {10.1561/3300000030},
	issn = {2474-1558},
	journal = {Foundations and Trends in Privacy and Security},
	number = {2–4},
	pages = {117-660},
	title = {Proofs, Arguments, and Zero-Knowledge},
	url = {http://dx.doi.org/10.1561/3300000030},
	volume = {4},
	year = {2022}
}

@book{AroraBarak2009, 
	author={Arora, Sanjeev and Barak, Boaz}, 
	title={Computational Complexity: A Modern Approach}, 
	publisher={Cambridge University Press}, 
	place={Cambridge}, 
	year={2009}
}

@Misc{CCKP19,
  author = {Chen, Shuo and Cheon, Jung Hee and Kim, Dongwoo and Park, Daejun},
  title        = {Verifiable Computing for Approximate Computation},
  howpublished = {{IACR} Cryptology ePrint Archive, Report 2019/762},
  year         = {2019},
  url          = {https://eprint.iacr.org/2019/762},
}

@article{Sumcheck-AFP-2024,
	author = {Garvía, Azucena and Christoph Sprenger and Jonathan Bootle},
	title = {The Sumcheck Protocol},
	journal = {Archive of Formal Proofs},
	issn = {2150-914x},
	note = {\url{https://isa-afp.org/entries/Sumcheck_Protocol.html}, Formal proof development},
	month = {feb},
	year = 2024
}


\begin{appendices}
\crefalias{section}{appsec}
\crefalias{subsection}{appsec}
\section{Completeness and Soundness of the Sumcheck Protocol}
\label{app:properties}

In this section, we carefully reprove the completeness and soundness properties of the sumcheck protocol (\Cref{prot:recursive_protocol}), thereby proving \Cref{thm:paper-sumcheck-theorem}.

In the soundness proof, we use probabilities in a loose manner without specifying the underlying distributions. The proof should be easy to follow without these details, which are explicitly defined in our formalization in \Cref{subsec:soundness}.

\subsection{Completeness}
\label{sapp:completeness_proof}

This proof is by induction on the arity of $\SPPolynomial$. We prove that \Cref{prot:recursive_protocol} will always accept an instance $\SPInstanceTuple \in \SPLanguage$. Recall that if $\SPInstanceTuple \in \SPLanguage$, then
\begin{equation}
\label{eq:sc-condition}
    \sum_{(\SumVal_1,\ldots,\SumVal_\SPArity) \in \SPSubset^\SPArity} \SPPolynomial(\SumVal_1,\ldots,\SumVal_\SPArity) = \SPTarget
    \enspace.
\end{equation}

\medskip
\emph{Base Case.} 
In the base case, $\SPPolynomial$ has arity 0, hence \eqref{eq:sc-condition} simplifies to $\SPPolynomial=\SPTarget$. The verifier $\Ver$ checks whether $\SPPolynomial = \SPTarget$, so the proof is accepted with probability $1$. This concludes the base case.

\medskip
\emph{Inductive Step.}
Suppose $p$ has arity $\SPArity>0$. The induction hypothesis states that for all $\NewSPInstanceTuple \in \SPLanguage$ where $\NewSPPolynomial$ has arity $\SPArity-1$, the verifier accepts with probability~$1$. 
We prove that the verifier accepts $\SPInstanceTuple \in \SPLanguage$ with probability~$1$.

Consider an execution of \Cref{prot:recursive_protocol}. The prover $\Pro$ computes the honest message
\begin{equation}
\label{eq:sc-honest-message}
    \SPMessagePolynomial(\XVar) \defeq \sum_{(\SumVal_2, \ldots, \SumVal_\SPArity) \in \SPSubset^{\SPArity-1}}  \SPPolynomial(\XVar, \SumVal_{2},\ldots, \SumVal_\SPArity)
    \enspace,
\end{equation}
and sends it to the verifier $\Ver$. Then $\Ver$ checks that $\SPMessagePolynomial$ is univariate, which is evident, and that $\SPMessagePolynomial$ has degree at most $\deg(\SPPolynomial)$,  which follows from the fact that $\SPMessagePolynomial$ is a sum of evaluations of $\SPPolynomial$.

The verifier check $\SPTarget \CheckEq \sum_{\SumVal \in \SPSubset}\SPMessagePolynomial(\SumVal)$ is satisfied because
\begin{equation*}
     v = \sum_{(\SumVal_1, \ldots, \SumVal_\SPArity) \in \SPSubset^\SPArity} p(\SumVal_1, \SumVal_2, \ldots, \SumVal_\SPArity) = \sum_{x \in H} q(x)
     \enspace,
\end{equation*}
where the first equality follows from \eqref{eq:sc-condition}, and the second follows from \eqref{eq:sc-honest-message}.

Since the arity $\SPArity$ of $\SPPolynomial$ is greater than 0, the protocol is now repeated over the instance $\NewSPInstanceTuple \in \SPLanguage$ where $\NewSPPolynomial$ is the  $(\SPArity-1)$-variate polynomial defined by $\NewSPPolynomial(\XVar_2, \ldots, \XVar_\SPArity) \defeq \SPPolynomial(\VerRandomness_1, \XVar_2, \ldots, \XVar_\SPArity)$ and $\NewSPTarget \defeq \SPMessagePolynomial(\VerRandomness_1).$ In order to apply the induction hypothesis it remains to show that $\NewSPInstanceTuple \in \SPLanguage$, which holds since
\begin{align*}
    \NewSPTarget &= \SPMessagePolynomial(\VerRandomness_1)
    = \sum_{(\SumVal_2, \ldots, \SumVal_\SPArity) \in \SPSubset^{\SPArity-1}} \SPPolynomial(\VerRandomness_1, \SumVal_2, \ldots, \SumVal_\SPArity)
    \enspace,\\
    &= \sum_{(\SumVal_2, \ldots, \SumVal_\SPArity) \in \SPSubset^{\SPArity-1}} \NewSPPolynomial(\SumVal_2, \ldots, \SumVal_\SPArity).
\end{align*}
Hence, $\Ver$ accepts the smaller instance $\NewSPInstanceTuple$ with probability 1 and therefore also the original instance $\SPInstanceTuple$. \qed

\subsection{Soundness}
\label{sapp:soundness_proof}

The proof is again by induction on the arity of $\SPPolynomial$. We show that for all instances $\SPInstanceShort = \SPInstanceTuple \notin \Language_\SPSymbol$ and for all provers $\Mal{\Pro}$ the verifier $\Ver$ accepts with probability at most $\SPArity \SPDegree / |\F|$ where $\SPArity$ is the arity of $\SPPolynomial$ and $\deg(\SPPolynomial) \leq \SPDegree$. 
Somewhat more formally, we show that
\begin{align*}
\Prob_{\SPRandomness_1, \ldots, \SPRandomness_n}[\Accept_{\SPInstanceShort}] & \leq \frac{\SPArity\SPDegree}{|\F|} \enspace,
\end{align*}
where $\Accept_{\SPInstanceShort} \defeq \langle \Mal{\Pro}, \Ver(\VerRandomness_1,\ldots,\VerRandomness_n) \rangle (\SPInstanceShort) = 1 \land \SPInstanceShort \notin \SPLanguage$, i.e., the verifier accepts the instance $\SPInstanceShort \notin \SPLanguage$.

\medskip
\emph{Base Case.} 
Recall that $\SPInstanceShort \notin \SPLanguage$ means
\begin{equation*}
\label{eq:sc-non-condition}
    \SPTarget \neq \sum_{(\SumVal_1, \ldots, \SumVal_\SPArity) \in \SPSubset^\SPArity} \SPPolynomial(\SumVal_1, \SumVal_2, \ldots, \SumVal_\SPArity)
    \enspace.
\end{equation*}
In the base case $\SPPolynomial$ has arity 0, hence this disequality simplifies to $\SPPolynomial \neq \SPTarget$. The verifier $\Ver$ checks whether $\SPPolynomial = \SPTarget$, so the proof is accepted with probability $0$. 

\medskip
\emph{Inductive Step.}
Suppose that $p$ has arity $\SPArity>0$. The induction hypothesis states that for all $\NewSPInstanceShort = \NewSPInstanceTuple \notin \SPLanguage$ where $\NewSPPolynomial$ has arity $\SPArity - 1$ and $\deg(\NewSPPolynomial) \leq d$, 
\begin{align*}
\Prob_{\SPRandomness_2, \ldots, \SPRandomness_n}[\Accept_{\NewSPInstanceShort}] & \leq \frac{(\SPArity-1)\SPDegree}{|\F|} \enspace.
\end{align*}
where $\Accept_{\NewSPInstanceShort} \defeq \langle \Mal{\Pro}, \Ver(\VerRandomness_2,\ldots,\VerRandomness_n) \rangle (\NewSPInstanceShort) = 1 \land \NewSPInstanceShort \notin \SPLanguage$.

We prove that the probability that $\Ver$ accepts the original instance $\SPInstanceShort \notin \SPLanguage$ is at most $\SPArity\SPDegree / |\F|$. Recall that the message $\SPMessagePolynomial$ sent by the honest prover $\Pro$ is defined by $\SPMessagePolynomial(\XVar) \defeq \sum_{(\SumVal_2, \ldots, \SumVal_\SPArity) \in \SPSubset^{\SPArity-1}}  \SPPolynomial(\XVar, \SumVal_{2}\ldots, \SumVal_\SPArity)$.
Suppose that $\Mal{\Pro}$ sends a message $\Mal{\SPMessagePolynomial}$ to the verifier. Since either $\PolyEq$ or $\PolyNeq$, we have
\begin{align*}
    \Prob[\Accept_{\SPInstanceShort}] & = 
    \Prob [\Accept_{\SPInstanceShort} \land \PolyEq] + 
    \Prob [\Accept_{\SPInstanceShort} \land \PolyNeq]
    \enspace.
\end{align*}
Note that $\Accept_{\SPInstanceShort}$ implies that $\Mal{\SPMessagePolynomial}$ is an univariate polynomial of degree at most $\SPDegree$ satisfying $\SPTarget = \sum_{\SumVal \in \SPSubset} \Mal{\SPMessagePolynomial}(\SumVal)$.

If $\PolyEq$, then
\begin{align*}
    \SPTarget &\neq \sum_{(\SumVal_1, \ldots, \SumVal_\SPArity) \in \SPSubset^\SPArity} \SPPolynomial(\SumVal_1, \SumVal_2, \ldots, \SumVal_\SPArity)
    \enspace,\\
    &= \sum_{\SumVal \in \SPSubset} \SPMessagePolynomial(\SumVal)
    = \sum_{\SumVal \in \SPSubset} \Mal{\SPMessagePolynomial}(\SumVal)
    \enspace,
\end{align*}
which contradicts 
$\Accept_{\SPInstanceShort}$.
Therefore, $\Prob [\Accept_{\SPInstanceShort} \land \PolyEq] = 0$ and 
\begin{align}
\label{eq:pr-acc-non-trivial}
\Prob[\Accept_{\SPInstanceShort}] = \Prob [\Accept_{\SPInstanceShort} \land \PolyNeq] \enspace.
\end{align}

Suppose that $\PolyNeq$. We make a further case split, on whether or not $\EvalEq$ holds:
\begin{align}
\label{eq:2nd-case-split}
  & \Prob [\Accept_{\SPInstanceShort} \land \SPMessagePolynomial \neq \Mal{\SPMessagePolynomial}] \nonumber \\
  & = \Prob [\Accept_{\SPInstanceShort} \land \PolyNeq \land \EvalEq ] \\
  & \;+ \Prob [\Accept_{\SPInstanceShort} \land \PolyNeq \land \EvalNeq ] \nonumber 
  \enspace.
\end{align}

We will now consider each of the two cases in turn. Note that $\EvalEq$ for at most $\SPDegree$ values of $\VerRandomness_1 \in \F$, since $\SPMessagePolynomial-\Mal{\SPMessagePolynomial}$ is a non-zero univariate polynomial of degree at most $\SPDegree$. Hence, we have
\begin{align}
\label{eq:roots-case}
 \Prob [\Accept_{\SPInstanceShort} \land \PolyNeq \land \EvalEq ] \leq \frac{n}{|\F|}
\end{align}

Forthe second probability in \eqref{eq:2nd-case-split}, we have
\begin{align}
& \Prob [\Accept_{\SPInstanceShort} \land \PolyNeq \land \EvalNeq ]
\nonumber \\
& \leq \Prob [\Accept_{\NewSPInstanceShort} ]
\label{eq:rec-case-1} \\
& \leq \frac{(\SPArity-1)\SPDegree}{|\F|} \enspace.
\label{eq:rec-case-2}
\end{align}

To see that the inequality \eqref{eq:rec-case-1} holds, consider the reduced instance $\NewSPInstanceShort = \NewSPInstanceTuple$, where $\NewSPPolynomial$ is the  $(\SPArity-1)$-variate polynomial 
$
\NewSPPolynomial(\XVar_2, \ldots, \XVar_\SPArity) \defeq \SPPolynomial(\VerRandomness_1, \XVar_2, \ldots, \XVar_\SPArity)
$
and $\NewSPTarget \defeq \Mal{\SPMessagePolynomial}(\VerRandomness_1)$. 
Since $\NewSPTarget = \Mal{\SPMessagePolynomial}(\VerRandomness_1) \neq {\SPMessagePolynomial}(\VerRandomness_1)$ and
\begin{align*}
    \SPMessagePolynomial(\VerRandomness_1)
    &= \sum_{(\SumVal_2, \ldots, \SumVal_\SPArity) \in \SPSubset^{\SPArity-1}} \SPPolynomial(\VerRandomness_1, \SumVal_2, \ldots, \SumVal_\SPArity)
    \enspace,\\
    &= \sum_{(\SumVal_2, \ldots, \SumVal_\SPArity) \in \SPSubset^{\SPArity-1}} \NewSPPolynomial(\SumVal_2, \ldots, \SumVal_\SPArity)
    \enspace,
\end{align*}
we have $\NewSPInstanceShort \notin\SPLanguage$. Hence, $\Accept_{\SPInstanceShort}$ implies $\Accept_{\NewSPInstanceShort}$, i.e., that the verifier $\Ver$ accepts $\NewSPInstanceShort\notin\SPLanguage$.
Finally, since $\NewSPPolynomial$ has arity $\SPArity-1$ and degree at most $\SPDegree$, we can justify \eqref{eq:rec-case-2} using the induction hypothesis.

Putting \eqref{eq:pr-acc-non-trivial}--\eqref{eq:rec-case-2} together, we obtain 
\begin{equation*}
    \Prob[\Accept_{\SPInstanceShort}] \leq \frac{\SPDegree}{|\F|} +  \frac{(\SPArity-1)\SPDegree}{|\F|} = \frac{\SPArity\SPDegree}{|\F|}
    \enspace.
\end{equation*}
which completes induction step and the soundness proof. \qed

\section{Univariate Polynomials and their Connection to Multivariate Polynomials}
\label{app:univariate-polys}

\subsection{Univariate Polynomials}
We use the univariate polynomials from Isabelle's polynomial theory \cite{IsabellePolynomial}, which formalizes the algebraic properties and operations of univariate polynomials. In this section we describe the definitions and lemmas taken from this theory which are needed in \Cref{app:roots_bound} to prove the roots bound.

The type of univariate polynomials with coefficients of type $\isa{'a}$ is defined by, 
$$
\isatypedef \isa{'a poly} =  \{\isa{f} \isa{::nat} \fun (\isa{'a::zero}). \:  \forall_{\infty} \:\isa{n}. ~ \isa{f} ~ \isa{n} = 0\}.
$$ 
The functions in this type map natural numbers, which correspond to the exponent of the monomials, to their coefficients in a type $\isa{'a}$ of typeclass $\isa{zero}$, which is the class of types with a $0$ element. This type can only contain functions which are eventually zero, meaning that only finitely many coefficients can be non-zero in a polynomial. Note that this type is isomorphic to the instance $\isa{nat} \fun_0 \isa{'a}$ of the type polynomial mappings introduced in \cref{sub:multivariate_polynomials}.
For example, informally speaking the polynomial $1 + 2x + 4x^3$ would be represented by the mapping $\{(0,1), (1,2), (3,4)\}$, sending all remaining numbers to 0. 

Equality of polynomials is expressed through the following lemma,
\begin{align*}
    \isabf{le}&\isabf{mma } \isa{poly\_eq\_iff: } \isa{p} = \isa{q} \longleftrightarrow (\forall \isa{n}. \: \isaco{coeff} ~ \isa{p} ~ \isa{n} = \isaco{coeff} ~ \isa{q} ~ \isa{n})
\end{align*}
where $\isaco{coeff}$ is the representation morphism of the type \isa{'a poly}, which takes a polynomial and a natural number $\isa{n}$ and returns the coefficient of the $\isa{n}^{\text{th}}$ power of the variable in the polynomial. The lemma states that $\isa{p}$ and $\isa{q}$ are equal if and only if their corresponding coefficients are equal for all powers of the variable.

\paragraph{Degree} The theory defines the degree of a polynomial by,
\begin{align*}
    & \isadefinition \isaco{degree} :: \isa{'a poly} \fun \isa{nat} \isawhere \\
    & \quad \isaco{degree} ~ \isa{p} = (\isaco{LEAST} ~ \isa{n}. \: \forall \isa{i}>\isa{n}.\: \isaco{coeff} ~ \isa{p} ~ \isa{i} = 0)
\end{align*}
so that the degree of a polynomial is the least exponent after which all coefficients are 0. We use the following two lemmas about the degree of a polynomial, 
\begin{align*}
    & \isalemma \isa{degree\_mult\_eq}: \\
    & \quad \isa{p} \neq 0 \Longrightarrow \isa{q} \neq 0 \Longrightarrow \isaco{degree} ~ (\isa{p}*\isa{q}) = \isaco{degree} ~ \isa{p} + \isaco{degree} ~ \isa{q}, \\
    & \isalemma \isa{degree\_power\_eq}: \\
    & \quad \isa{p} \neq 0 \Longrightarrow \isaco{degree} ~ \isa{p}^\isa{n} = \isa{n} \cdot \isaco{degree} ~ \isa{p}.
\end{align*}

\paragraph{Polynomial roots} The identifier \isa{poly} is overloaded and refers both to the type of univariate polynomials and their evaluation function; the expression $\isaco{poly} ~ \isa{p} ~ \isa{a}$ evaluates the polynomial $\isa{p}$ at the value $\isa{a}$. Therefore  the following lemma states the finiteness of the roots of a polynomial,
\begin{align*}
    \isabf{le}&\isabf{mma } \isa{poly\_roots\_finite: } \isa{p} \neq 0 \Longrightarrow \isaco{finite} ~ \{\isa{a} . \: \isaco{poly} ~ \isa{p} ~ \isa{a} = 0\}.
\end{align*}

\paragraph{Order of polynomial roots} The order of an element in a univariate polynomial is defined as follows,
\begin{align*}
    & \isadefinition \isaco{order} :: \isa{'a} \fun \isa{'a poly} \fun \isa{nat} \isawhere \\
    & \quad \isaco{order} ~ \isa{a} ~ \isa{p} = (\isaco{LEAST} ~ \isa{n}. \: \neg \: ([:-\isa{a}, 1:]^{\isaco{Suc} ~ \isa{n}} ~ \isaco{dvd} ~ \isa{p}
    ))
\end{align*}
where $[:-\isa{a}, 1:]$ represents the polynomial $(\isa{x}-\isa{a})$, supposing the only variable of $\isa{p}$ is denoted by $\isa{x}$. The order of $\isa{a}$ in $\isa{p}$ is therefore the maximum number of times which $(\isa{x}-{a})$ divides the polynomial.

There are two useful lemmas regarding the order function,
\begin{align*}
    & \isalemma \isa{order\_root}: \\ 
    & \quad\isaco{poly} ~ \isa{p} ~ \isa{a} = 0 \:\longleftrightarrow\: \isa{p}=0 \: \lor \: \isaco{order} ~ \isa{a} ~ \isa{p} \neq 0 \isa{,}\\[1ex]
    & \isalemma \isa{order\_decomp}: \\
    & \quad\isaassumes \isa{p} \neq 0 \\
    & \quad\isashows \exists \isa{q}. \; \isa{p} = [:-\isa{a},1:]^{\isaco{order} ~ \isa{a} ~ \isa{p}} \cdot \isa{q} \\
    & \phantom{\quad\isashows \exists \isa{q}. }
      \isand \neg \: ([:-\isa{a}, 1:] ~ \isaco{dvd} ~ \isa{q})
\end{align*}

\subsection{Connecting Univariate and Multivariate Polynomials}

The theory \textit{MPoly\_Type\_Univariate} of the executable multivariate polynomials library~\cite{Polynomials-AFP} connects the univariate polynomials described above to the multivariate polynomials from \cref{sub:multivariate_polynomials}. 
More specifically, it defines two mutually inverse morphisms $\isaco{poly\_to\_mpoly}$ and $\isaco{mpoly\_to\_poly}$, which connect polynomials of type $\isa{'a poly}$ to multivariate polynomials of type $\isa{'a mpoly}$ and vice versa. These morphisms will allow us to transfer theorems about univariate polynomials to univariate multivariate polynomials in our proof of the roots assumption (\Cref{app:roots_bound}).

We highlight one lemma which will be needed in our final proof of the roots bound. This lemma states that the evaluation functions $\isa{poly}$ and $\isa{insertion}$ of univariate and multivariate polynomials, respectively, are equivalent for univariate multivariate polynomials. It is,
\begin{align*}
    & \isalemma \isa{poly\_eq\_insertion}: \\
    & \quad \isaassumes \isaco{vars} ~ \isa{p} \subseteq \{\isa{v}\}\\
    & \quad \isashows \isaco{poly} ~ (\isaco{mpoly\_to\_poly} ~ \isa{v} ~ \isa{p}) ~ \isa{x} = \isaco{insertion} ~ (\lambda \isa{v}.\: \isa{x}) ~ \isa{p}
\end{align*}

\section{Roots Bound for Univariate and Multivariate Polynomials}
\label{app:roots_bound}

In this section, we show that the number of roots of a univariate polynomial $\isa{p}$ of type $\isa{'a mpoly}$ is bounded by its total degree $\degco ~ \isa{p}$.
We refer to the degree of a univariate polynomial $\isa{p}$ as $\isaco{udegree} ~ \isa{p}$ and the degree of a variable $\isa{v}$ in a multivariate polynomial $\isa{p}$ as $\isaco{degree} ~ \isa{p} ~ \isa{v}$.

We first establish the following roots bound for univariate polynomials of type $\isa{'a poly}$:
\begin{align*}
    & \isalemma \isa{univariate\_roots\_bound}: \\
    & \quad \isaassumes \isa{p} \neq 0 \\
    & \quad \isashows \isaco{card} ~ \{\isa{x} :: \isa{'a}. \; \isaco{poly} ~ \isa{p} ~ \isa{x} = 0\} \leq \isaco{udegree} ~ \isa{p}.
\end{align*}

\begin{proof}[Proof (sketch)]
We proceed by strong induction on the degree of the polynomial $\isa{p}$, using some lemmas from Appendix~\ref{app:univariate-polys}. We find a root $\isa{r}$ of $\isa{p}$ (if there are none then the proof is trivial) and write $\isa{p} = (\isa{x}-\isa{r}) \cdot \isa{q}$ for some $\isa{q}$ using the lemma \isa{order\_decomp}.  By lemma \isa{degree\_mult\_eq} we derive that the degree of $\isa{p}$ is strictly larger than that of $\isa{q}$ and therefore we can apply the inductive hypothesis to conclude that $\isa{q}$ has strictly less roots than the degree of $\isa{p}$. Since $(\isa{x}-\isa{r})$ has exactly one root, and since we can show that the roots of $\isa{p}$ are less than or equal to the roots of $(\isa{x}-\isa{r})$ unioned to those of $\isa{q}$, we can conclude that $\isa{p}$ has at most $\isaco{udegree} ~ \isa{p}$ roots. This last step uses the lemmas \isa{order\_root} and \isa{degree\_power\_eq}. 
\end{proof}

In order to transfer this result to univariate polynomials of type $\isa{'a mpoly}$, we prove two lemmas relating the different degree functions $\isaco{udegree}$, $\isaco{degree}$ and $\degco$. These are 
\begin{align*}
    & \isalemma \isa{udegree\_eq\_degree}:\\
    & \quad \isaassumes \isaco{vars} ~ \isa{p} \subseteq \{\isa{v}\} \\
    & \quad \isashows \isaco{udegree} ~ (\isaco{mpoly\_to\_poly} ~ \isa{v} ~ \isa{p} ) = \isaco{degree} ~ \isa{p} ~ \isa{v} \\[1ex]
    & \isalemma \isa{degree\_eq\_total\_degree}: \\
    & \quad \isaassumes \isaco{vars} ~ \isa{p} \subseteq \{\isa{v}\} \\
    & \quad \isashows \degco ~ \isa{p} = \isaco{degree} ~ \isa{p} ~ \isa{v}
\end{align*}
Their proofs rely on basic properties of polynomials, definitions of the degree functions, and properties of the $\isaco{Max}$ and $\isaco{LEAST}$ functions. 

Finally, using the lemmas proved above and the lemma \isa{poly\_eq\_insertion} of Appendix~\ref{app:univariate-polys}, it is straightforward to prove the following roots bound:
\begin{align*}
    & \isalemma \isa{univariate\_mpoly\_roots\_bound:}\\
    & \quad \isaassumes \isaco{vars} ~ \isa{p} \subseteq \{\isa{v}\} \isaand \isa{p} \neq 0\\
    & \quad \isashows \isaco{card} ~ \{\isa{x}. \: \isaco{insertion} ~ (\lambda \isa{v} . \: \isa{x}) ~  \isa{p} = 0\} \leq \degco ~ \isa{p}.
\end{align*}

Note that $\isa{roots}$ assumption from \Cref{fig:locale} differs from the statement of this lemma in that it refers to two polynomials instead of one. However, it is easy to prove the $\isa{roots}$ assumption from the lemma above by applying it to the difference of polynomials $\isa{p} - \isa{q}$, since $\isaco{eval}~(\isa{p} - \isa{q})~[ x \mapsto r] = 0$ implies that $\isaco{eval}~\isa{p}~[x \mapsto r] = \isaco{eval}~\isa{q}~[x \mapsto r]$.

\end{appendices}


\end{document}